\documentclass[11pt, oneside]{article}   	
\usepackage{slashed}
\usepackage{jheppub}
\usepackage{amsmath}
\usepackage{amssymb}
\usepackage{amsfonts}
\usepackage{amsthm}
\usepackage{amsbsy}
\usepackage{array}
\usepackage{mathtools}
\usepackage[usenames,dvipsnames]{xcolor}
\usepackage{subcaption}
\usepackage{shorthand}
\usepackage{graphicx}
\usepackage{comment}
\usepackage[vcentermath]{youngtab}
\usepackage{hyperref}
\usepackage{cleveref}
\usepackage{cancel}
\usepackage[normalem]{ulem}
\usepackage{tikz}
\usetikzlibrary{snakes}
\usetikzlibrary{decorations}
\usetikzlibrary{trees}
\usetikzlibrary{decorations.pathmorphing}
\usetikzlibrary{decorations.markings}
\usetikzlibrary{external}
\usetikzlibrary{intersections}
\usetikzlibrary{shapes,arrows}
\usetikzlibrary{arrows.meta}
\usetikzlibrary{calc}
\usetikzlibrary{shapes.misc}
\usetikzlibrary{decorations.text}
\usetikzlibrary{backgrounds}
\usetikzlibrary{fadings}
\usetikzlibrary{tikzmark,calc,arrows,shapes,decorations.pathreplacing}
\usetikzlibrary{patterns}
\tikzset{
        cross/.style={cross out, draw=black, minimum size=2*(#1-\pgflinewidth), inner sep=0pt, outer sep=0pt},
	branchCut/.style={postaction={decorate},
		snake=zigzag,
		decoration = {snake=zigzag,segment length = 2mm, amplitude = 2mm}	
    }}
\newcommand{\bea}{\setlength\arraycolsep{2pt} \begin{eqnarray}}
\newcommand{\eea}{\end{eqnarray}}

\def\fft#1#2{{\frac{#1}{#2}}}

\newcommand{\baa}{\begin{align}}
\newcommand{\eaa}{\end{align}}

\makeatletter
\def\@fpheader{\ }
\makeatother

\title{%Bouncing singularities and defect conformal field theories on thermal Wilson lines 
Bouncing singularities and thermal correlators on line defects
}
\author{Simone Giombi$^1$, Yue-Zhou Li$^1$, Jieru Shan$^1$}
\affiliation{
${}^1$ Department of Physics, Princeton University, Princeton, NJ 08544, USA \\
}
\emailAdd{sgiombi@gmail.com}
\emailAdd{liyuezhou@princeton.edu} 
\emailAdd{jierus@princeton.edu}
\date{}
\abstract{Thermal correlators in holographic conformal field theories are known to exhibit singularities in complex time, sometimes referred to as ``bouncing singularities", which are believed to be related to bulk geodesics probing the black hole interior. These singularities correspond to exponentially suppressed contributions in the high-frequency limit of the thermal correlators. We revisit in detail the calculation of retarded two-point functions of local operators dual to bulk scalar fields in the planar AdS black hole background. We confirm that these correlators develop bouncing singularities, and highlight the agreement of two independent methods: a large frequency WKB analysis with infalling boundary conditions at the horizon; and an asymptotic OPE analysis that relies only on the near-boundary expansion, without any direct input from the black hole interior. We then extend these calculations to the case of the retarded two-point function of displacement operators on a Wilson line in the finite temperature gauge theory. This is computed holographically by solving the wave equation for the transverse fluctuations of the dual string worldsheet in the planar AdS black hole background. We find that these defect correlators also exhibit bouncing singularities, and again observe exact agreement between the WKB analysis sensitive to the black hole interior and the asymptotic OPE analysis. This agreement suggests that the bouncing singularities and the corresponding OPE data encode a universal high-frequency structure of the retarded correlators, and we propose a factorization formula that encodes the deviations from this universality.}

\begin{document}

\maketitle
\pagenumbering{roman}
\setcounter{page}{2}
\newpage
\pagenumbering{arabic}
\setcounter{page}{1}

\section{Introduction}

The interior of a black hole, particularly the singularity where classical spacetime becomes incomplete, is among the most mysterious structures in gravitational physics and may offer deep insights into quantum gravity. A central question is whether and how the black hole interior can be probed from the outside, especially by asymptotic observers.

More broadly, this can be formulated in a sharper way: from the perspective of asymptotic observers, how can one probe the internal structure of a gravitational source and determine whether it is truly a black hole or instead some other compact object? This question has become one of the central agendas in the era of gravitational wave astronomy, following the detections by LIGO/Virgo/KAGRA \cite{LIGOScientific:2014pky,VIRGO:2014yos,KAGRA:2020agh}. Through gravitational wave observations, we expect to extract detailed information about black holes, neutron stars, and more general compact binaries from asymptotic correlation functions of gravitational radiation.

Holography \cite{Maldacena:1997re,Witten:1998qj,Gubser:1998bc} may put these questions on a firmer footing. Placing black holes and other compact objects in Anti de-Sitter space (AdS) and probing them with bulk waves in the classical limit essentially corresponds to computing correlation functions in nontrivial states of the conformal field theory (CFT) in the large-$N$ limit \cite{Freedman:1998tz,Son:2002sd,Policastro:2001yc,Policastro:2002se,Skenderis:2008dh,Skenderis:2008dg}. The problem then becomes to understand which behaviors of CFT correlators capture the fine structure of black holes and compact objects, namely to identify which parts are universally dictated by the underlying dynamics and which parts are genuinely sensitive to the specific state being probed.

In particular, when probing black holes, the boundary CFT is in a thermal ensemble \cite{Maldacena:2001kr}. Remarkably, it has been shown that the geodesic length and proper time associated with geodesics approaching the black hole singularity are encoded in thermal CFT correlators \cite{
Fidkowski:2003nf,Festuccia:2005pi,Festuccia:2008zx,Amado:2008hw,Grinberg:2020fdj,David:2022nfn,Ceplak:2024bja,Ceplak:2025dds,Afkhami-Jeddi:2025wra,Dodelson:2025jff,AliAhmad:2026wem}. The striking aspect is that although the black hole singularity itself is not a well-defined object within classical gravity, its imprint nevertheless appears in thermal correlators of the CFT. More specifically, the geodesic that ``bounces off" the singularity exhibits a complex time delay, which manifests itself as a singularity in thermal correlators at complex time $t=t_c=\beta/2(1+i)$, dubbed as the ``bouncing singularity". This structure appears either in two-sided Wightman functions of the thermofield double state \cite{Parisini:2023nbd,Ceplak:2024bja,Ceplak:2025dds,Dodelson:2025jff} or, as discussed more recently, in retarded correlators \cite{Afkhami-Jeddi:2025wra,Jia:2025jbi} (which are the most directly accessible classical observables). 

In recent years, deeper insights have emerged that go beyond the simple geometric picture relating the singularities of thermal correlators and geodesics that bounce off the black hole singularity. From the bulk perspective, it has been shown that these singularities are controlled by the asymptotic quasinormal modes of black holes \cite{Dodelson:2025jff}, via the thermal product formula \cite{Dodelson:2023vrw}. From the CFT perspective, it was recently pointed out that the bouncing singularities are governed by multi-stress-tensor operators $T^n$ in the thermal OPE in the asymptotic regime $n \to \infty$ \cite{Parisini:2023nbd,Ceplak:2024bja}. In parallel, these singularities are reflected in nonperturbative tails $\sim e^{-\beta\omega/2}$ of retarded correlators at large real frequency, which improve the asymptotic high-frequency expansion in powers of $1/\omega$. These exponentially small terms can be directly computed using a WKB-based approach \cite{Afkhami-Jeddi:2025wra}, which requires a careful analysis of the wave equation near the black hole singularity. This subject remains an active arena for uncovering deeper aspects of black hole physics. Recent developments include generalizations to charged black holes that probe naked singularities \cite{Ceplak:2025dds,Dodelson:2025jff}, the emergence of more intricate bouncing structures at large transverse momentum \cite{Jia:2025jbi}, connections to bulk-cone singularities \cite{Dodelson:2020lal,Dodelson:2023nnr} and eikonal scattering \cite{Kulaxizi:2017ixa,Kulaxizi:2018dxo,Kulaxizi:2019tkd,Parnachev:2020zbr,DiVecchia:2023frv,Araya:2026shz,Jia:2026pmv}, attempts on the generalization to higher-point \cite{Chakravarty:2025ncy}, and the perspectives from thermal conformal bootstrap \cite{Iliesiu:2018fao,Parnachev:2020fna,Alday:2020eua,Niarchos:2025cdg,Buric:2025anb,Buric:2025fye,Barrat:2025nvu,Barrat:2025twb}, among others.

Nonetheless, the aforementioned explorations have so far been restricted to correlation functions of local operators on the CFT side. More generally, one may also consider line defects in thermal CFTs, which naturally arise, for example, from heavy quarks propagating through the quark-gluon plasma \cite{Rey:1998ik}. A simple framework to model this setup is to consider Wilson lines in the CFT \cite{Maldacena:1998im,Erickson:2000af} and study how they are deformed or deflected by inserting displacement operators along the line and computing their correlation functions \cite{Drukker:2006xg,Cooke:2017qgm,Sakaguchi:2007ba,
Giombi:2017cqn,Beccaria:2017rbe,Giombi:2018qox,
Giombi:2018hsx,Beccaria:2019dws,Giombi:2020amn, Liendo:2018ukf, Ferrero:2021bsb,  Giombi:2022pas, Giombi:2023zte}, which is effectively described by a one dimensional defect CFT. Here we are interested in the correlators of this defect CFT at finite temperature (see \cite{Barrat:2024aoa} for recent work on line defects at finite temperature).  

In this paper, we address a natural question: can a Wilson line also probe bouncing singularities, manifested through nontrivial structures in the thermal correlators of the one dimensional defect CFT? The answer is yes, and we indeed find that the retarded correlator of displacement operators displays the same bouncing singularity as in previous discussions without defects, although with different details. Holographically, the thermal correlators can be obtained by solving the wave equation for the transverse fluctuations of the static string dual to a Wilson line along the time direction. The string worldsheet has an asymptotically AdS$_2$ geometry, with a horizon inherited from the horizon of the bulk black hole. We perform the calculation of the thermal defect correlators both using the WKB method of \cite{Afkhami-Jeddi:2025wra}, which directly probes the nonperturbative tails $G_R(\omega)\sim e^{-\beta/2(1-i)\omega}$, and the asymptotic OPE method of \cite{Fitzpatrick:2019zqz,Fitzpatrick:2019efk}, which solves the wave equation in a near-boundary expansion. From the OPE perspective, we find that the bouncing singularity is now controlled by multi-stress-tensor operators inserted along the line defect, schematically denoted as $W T^n$. We find that the predictions of the WKB analysis and the asymptotic OPE method precisely agree, even to subleading orders in $1/n$. In the bulk, our results suggest that the correlators of transverse fluctuations of the string can probe features of the interior, as encoded in the string worldsheet.  

%In the bulk, this implies that the transverse fluctuations of the string see the black hole singularity as encoded in the string worldsheet.

The agreement between the WKB analysis and the asymptotic OPE analysis was also found for ``bulk" thermal correlators \cite{Afkhami-Jeddi:2025wra}, and it persists along the Wilson line. At first sight, this is surprising, because the WKB method in \cite{Afkhami-Jeddi:2025wra} relies on the details near the black hole singularity, whereas the asymptotic OPE method \cite{Fitzpatrick:2019zqz,Fitzpatrick:2019efk} relies only on the near-boundary expansion. More precisely, the asymptotic OPE method, developed as a tool to study heavy-light-light-heavy correlators \cite{Fitzpatrick:2019zqz,Fitzpatrick:2019efk,Li:2019tpf,Li:2019zba,Karlsson:2019qfi,Karlsson:2019dbd,Karlsson:2019txu,Karlsson:2020ghx,Li:2020dqm,Fitzpatrick:2020yjb,Abajian:2023jye,Dodelson:2022eiz,Karlsson:2022osn,Dodelson:2022yvn,Huang:2024wbq}, does not know whether the deep interior contains a black hole or a compact heavy object; therefore, the results are valid for both thermal states and other heavy states. This strongly suggests that the bouncing singularity, although believed to be specifically relevant to the black hole singularity, is controlled by universal dynamics regardless of the specific state being probed. This is a reminder of the universality of thermalization in CFTs \cite{Srednicki:1999bhx,DAlessio:2015qtq,Lashkari:2016vgj,Collier:2019weq,Delacretaz:2020nit}, see for example \cite{Karlsson:2022osn,Karlsson:2021duj,Dodelson:2022yvn,Esper:2023jeq}. To understand this aspect more deeply, we thus propose a high-frequency factorization formula for retarded correlators to specify the universal piece and the state-dependent piece, and claim that the bouncing singularity is encoded in the universal piece that contains the multi-stress-tensor OPEs, although whether they can be properly resummed to signal the black hole singularity depends on the details of the state.

The rest of the paper is organized as follows. We first review the basics of thermal two-point functions in CFTs in section \ref{subsec: thermal correlator}. We then revisit the holographic thermal retarded correlators of local operators in section \ref{sec: thermal correlator bulk}, which includes details of the WKB proposal of \cite{Afkhami-Jeddi:2025wra} in section \ref{subsec: WKB bouncing}, and the asymptotic OPE method \cite{Fitzpatrick:2019zqz,Fitzpatrick:2019efk} in section \ref{sec: bouncing from OPE p0}. In addition, we extend the numerical asymptotic OPE analysis of \cite{Ceplak:2024bja} to higher orders in $1/n$ in section \ref{subsec: x0 comments}. In section \ref{sec: Wilson line}, we introduce the thermal Wilson line setup as a defect CFT, and review its holographic description in the bulk. In section \ref{sec: thermal Wilson line correlators}, we study the thermal retarded correlators along a line defect using holography in detail, including revisiting the low-frequency stochastic regime in section \ref{subsec: small frequency}, computing the bouncing singularity using the WKB method in section \ref{subsec: defect bouncing WKB}, and using the defect OPE in section \ref{subsec: defect OPE}. We also perform numerical checks of our analytic results in both the low-frequency and high-frequency regimes in section \ref{subsec: numerical}, and discuss the exact solution for the thermal correlator of massless modes on the string, which does not show the bouncing singularity, in section \ref{subsec: massless}. We discuss the universality and a proposed factorization formula in section \ref{sec: universality}, and conclude with a summary and future directions in section \ref{sec: summary}. We collect additional technical details in the Appendices.

\section{Thermal correlators in CFTs}
\label{subsec: thermal correlator}

Our goal is to understand two-point functions in thermal CFTs, such that the Euclidean theory is on the topology $S_\beta\times \mathbb{R}^{d-1}$. In Eucliden signature, the thermal two-point function, which is time-ordered in Euclidean time, is 
\be
G_E(\tau,x)=\langle \mathcal{O}(\tau,x)\mathcal{O}(0,0)\rangle_\beta:={\rm Tr}\left[e^{-\beta H}\mathcal{O}(\tau,x)\mathcal{O}(0,0)\right]\,. 
\ee
This satisfies the KMS condition $G_E(\tau,x)=G_E(\beta-\tau,0)$. To study the real time physics, we perform analytic continuation in complex time $\tau=\epsilon+i t$, which allows one to define the Wightman functions in real time
\be
G^>(t,x)=G_E(\epsilon+i t,x)\,,\quad G^<(t,x)=G_E(-\epsilon+i t,x)\,.
\ee
The retarded correlator that makes causality manifest is 
\be
G_R(t,x)=i\left(G^>(t,x)-G^<(t,x)\right) \theta(t)\,,
\ee
which satisfies the fluctuation-dissipation theorem (FDT) in frequency space
\be
G^{>}(\omega)= \fft{2{\rm Im}\, G_R(\omega)}{1-e^{-\beta \omega}}\,.
\ee

In CFTs, OPE is a well-posed organization principle, especially in Euclidean signature, which gives the Euclidean thermal block expansion \cite{Iliesiu:2018fao}
\be
G_E(\tau;x)=\sum_{\Delta',J} c_{\Delta',J}\fft{\hat{C}^{(\fft{d}{2}-1)}_J\left(\eta\right)}{\left(\tau^2+x^2\right)^{\Delta-\fft{\Delta'}{2}}}\langle \mathcal{O}'\rangle_{\beta}\,,\label{eq: block expansion T}
\ee
where $\eta=\tau/\sqrt{x^2+\tau^2}$ and $\langle\mathcal{O}'\rangle_{\beta}=\beta^{-\Delta'}$. The normalized Gegenbauer polynomial $\hat{C}_J$ is defined by
\be
\hat{C}_J^{\fft{d}{2}-1}(x)=\,_2F_1\left(-J,J+d-2,\fft{d-1}{2},\fft{1-x}{2}\right)\,,\quad \hat{C}_J^{\fft{d}{2}-1}(1)=1\,.
\ee
This Euclidean thermal block is largely simplified at $x=0$, where we have
\be
G_E(\tau,x=0)=\fft{1}{\tau^{2\Delta}}\sum_{\Delta'.J}c_{\Delta',J} \left(\fft{\tau}{\beta}\right)^{\Delta'}=\fft{1}{\tau^{2\Delta}}\sum_{\Delta'}a_{\Delta'} \left(\fft{\tau}{\beta}\right)^{\Delta'}\,.
\ee
We have defined the ``effective'' OPE coefficients
\be
a_{\Delta'}=\sum_{J:\Delta'_{\rm phy}(J)=\Delta'}c_{\Delta',J}\,.\label{eq: OPE x0}
\ee
In this case,  the analytic continuation can be easily implemented
\be
G^>(t,0)=t^{-2\Delta}\sum_{\Delta'} a_{\Delta'} e^{i\fft{\pi}{2}\left(\Delta'-2\Delta\right)}\left(\fft{ t}{\beta}\right)^{\Delta'}\,.
\ee
Therefore, we have the OPE for retarded Green's function with $x=0$ as
\be
G_R(t,0)=-2 t^{-2\Delta}\sum_{\Delta'} a_{\Delta'}\sin\left(\fft{\Delta'-2\Delta}{2}\pi\right)\left(\fft{ t}{\beta}\right)^{\Delta'}\,.
\ee

It is also insightful to consider the momentum space, which is more convenient in some holographic calculations. We can perform the Fourier transform as
\be
G_E(\tau,p)=\int d^{d-1}x G_E(\tau,x)e^{-ip\cdot x}\,.
\ee
We thus find (take $\beta=1$ for simplicity)
\be
&G_E(\tau,p)=\sum_{\Delta',J}c_{\Delta',J}\sum_{k=0}^{\fft{J}{2}} \alpha_{k,J} \tau^{J-2k} \left(\fft{2\tau}{p}\right)^{\fft{\Delta'-J-2\Delta+2k+d-1}{2}}K_{\fft{d-2+\Delta'-J-2\Delta+2k}{2}}(p \tau)\,,\nn\\
& \alpha_{k,J}=\frac{(-1)^k 2^{d-2+J-2k}\pi^{\fft{d}{2}-1} \Gamma \left(\frac{d-1}{2}\right) \Gamma (J+1) \Gamma \left(\frac{d}{2}+J-k-1\right)}{\Gamma (k+1) \Gamma (d+J-2) \Gamma (J-2 k+1) \Gamma \left(\frac{1}{2} \left(J-2 k-\Delta' +2 \Delta\right)\right)}\,,\label{eq: G_E p space}
\ee
which agrees with \cite{Manenti:2019wxs}. We can thus take zero momentum $p=0$, which simplifies the thermal block expansion
\be
G_E(\tau,p=0)=\tau^{-2\Delta+d-1}\sum_{\Delta'}\hat{a}_{\Delta'} \left(\fft{\tau}{\beta}\right)^{\Delta'}\,,\label{eq: G_E p0}
\ee
where
\be
\hat{a}_{\Delta'}=\sum_{J: \Delta'_{\rm phy}(J)=\Delta'} \frac{\pi ^{d/2} 2^{d+\Delta' -2 \Delta _{\mathcal{O}}} \Gamma \left(-d-\Delta' +2 \Delta _\mathcal{O}+1\right)}{\Gamma \left(\frac{J}{2}+\Delta _\mathcal{O}-\frac{\Delta' }{2}\right) \Gamma \left(-\frac{d}{2}+\Delta _\mathcal{O}-\frac{J}{2}-\frac{\Delta' }{2}+1\right)} c_{\Delta',J}\,.\label{eq: a def}
\ee
The corresponding retarded correlator has the expansion
\be
G_R(t,p=0)&=2t^{d-1-2\Delta_{\mathcal{O}}}\sum_{\Delta'} \hat{a}_{\Delta'} \cos\left(\fft{d+\Delta'-2\Delta}{2}\pi\right)\left(\fft{t}{\beta}\right)^{\Delta'}\nn\\
& :=t^{d-1-2\Delta_{\mathcal{O}}}\sum_{\Delta'} \hat{a}_{\Delta'}^{Rt} \left(\fft{t}{\beta}\right)^{\Delta'}\,,\label{eq: GR t OPE}
\ee

It is often convenient to work in the frequency space, where the fluctuation-dissipation theorem is manifest. By explicitly performing the Fourier transform assuming even integer spins, we find (this agrees with \cite{Dodelson:2022eiz})
\be
& G^>(\omega,p)=\theta(\omega)\theta(\omega^2-p^2)\sum_{\Delta',J}c_{\Delta',J}\hat{\alpha}_{\Delta',J}e^{i\fft{\Delta'-2\Delta+d-1}{2}\pi} \Gamma\left(d+\Delta'-2\Delta\right) \left(-i\omega\right)^{-d-\Delta'+2\Delta_0}\times\nn\\
& \left(1-\xi^2\right)^{\fft{-d-J-\Delta'+2\Delta}{2}}\,_2F_1\left(-\fft{J}{2},\fft{1-J}{2},\fft{d-1}{2},\xi^2\right)\,,\label{eq: GR OPE p and omega}
\ee
where $\xi=p/\omega$. Thus we have (keeping the theta function implicit)
\be
& G_R(\omega,p)=2\sum_{\Delta',J}\hat{\alpha}_{\Delta',J} \cos\left(\fft{d+\Delta'-2\Delta}{2}\pi\right)\Gamma\left(d+\Delta'-2\Delta\right) \left(-i\omega\right)^{-d-\Delta'+2\Delta}\times\nn\\
& \left(1-\xi^2\right)^{\fft{-d-J-\Delta'+2\Delta}{2}}\,_2F_1\left(-\fft{J}{2},\fft{1-J}{2},\fft{d-1}{2},\xi^2\right)\,,
\ee
We can double check this result by taking $p=0$ and compare with the simple Fourier transform of \eqref{eq: GR t OPE}
\be
G_R(\omega,p=0)&=2\left(-i\omega\right)^{2\Delta-d}\sum_{\Delta'} \hat{a}_{\Delta'} \cos\left(\fft{d+\Delta'-2\Delta}{2}\pi\right) \Gamma\left(d+\Delta'-2\Delta\right) \left(-i\beta \omega\right)^{-\Delta'}\nn\\
& := \left(-i\omega\right)^{2\Delta-d}\sum_{\Delta'} \hat{a}_{\Delta'}^{R\omega}  \left(-i\beta \omega\right)^{-\Delta'}\,.\label{eq: GR OPE}
\ee

Some comments are in order. Strictly speaking, the Euclidean OPE \eqref{eq: block expansion T} is valid in the regime $\tau^2+x^2<\beta^2$, while the Fourier transform in both space and time integrates over regions beyond this domain of convergence. Therefore, the resulting momentum and frequency space OPEs, such as \eqref{eq: G_E p space} and \eqref{eq: GR OPE p and omega}, should really be understood as asymptotic expansions: in the large momentum or high frequency limit, the transform is asymptotically controlled by short distances, where the Euclidean OPE remains valid, while long distance data give contact terms or nonperturbative contributions beyond the power law OPE. Related discussions can be found in \cite{Caron-Huot:2009ypo} and \cite{Manenti:2019wxs}.

In this paper, we also study thermal two-point functions on line defects, which exhibit essentially the same structure as the $x=0$ case for bulk scalars, since they can be understood as correlators in 1D defect CFT. Namely, we can write
\be
&G_{E}^{\rm defect}(\tau)=\fft{1}{\tau^{2\Delta}}\sum_{\Delta'}a_{\Delta'} \left(\fft{\tau}{\beta}\right)^{\Delta'}\,,\nn\\
& G_R^{\rm defect}(t)=-2 t^{-2\Delta}\sum_{\Delta'} a_{\Delta'}\sin\left(\fft{\Delta'-2\Delta}{2}\pi\right)\left(\fft{ t}{\beta}\right)^{\Delta'}\,.
\label{GR-defect-OPE}
\ee

\section{Thermal retarded correlators in holography}
\label{sec: thermal correlator bulk}

In this section, we revisit the computations of retarded correlators in holographic CFTs at finite temperature, setting up the technical groundwork.

\subsection{Holographic set-up}

Our goal is to study the retarded correlator in thermal holographic CFTs at large $N$, focusing on the large frequency behavior and its relation to the OPE coefficients $\hat{a}_{\Delta'}^R$. In holography, the thermal two-point function can be computed by studying linearized bulk fields propagating in an asymptotically AdS planar black hole background
\be
ds^2=-\fft{r^2 f(r)}{R_{\rm AdS}^2} dt^2 + \fft{R_{\rm AdS}^2}{r^2 f(r)}dr^2 +\fft{r^2}{R_{\rm AdS}^2} dx^2\,,\quad f(r)=1-\fft{\mu}{r^{d}}\,,\label{eq: bh metric}
\ee
where
\be
\mu=r_h^{d}= \left(\fft{4\pi R_{\rm AdS}^2}{\beta d}\right)^d= \fft{16 \pi G M R_{\rm AdS}^2}{d-1}=\fft{4\Gamma\left(d+2\right)}{(d-1)^2 \Gamma\left(\fft{d}{2}\right)^2} \fft{\Delta_H}{C_T}R_{\rm AdS}^{d}\,.
\ee
The black hole corresponds to a thermal ensemble in CFT, generated by thermalization of heavy states with dimensions $\Delta_H \sim M R_{\rm AdS}$. For simplicity, we take $R_{\rm AdS}=1$ and we work only in $d=4$. For convenience, we may also take $\mu=r_h=1$, such that $\beta=\pi$ in $d=4$. The temperature dependence can be easily restored by dimensional analysis.

To compute the thermal retarded correlator of a scalar primary operator with scaling dimension $\Delta$, we consider a scalar field minimally coupled to gravity in AdS
\be
S_{\phi}=-\fft{1}{2} \int d^{d+1}x\sqrt{-g}\left((\nabla\phi)^2+ m^2 \phi^2\right)\,,\quad m^2=\Delta(\Delta-d)\,.
\ee
The equation of motion is
\be
r^{3} \left(\left((d+1)f+r f'\right)\partial_r \phi+r \partial_r^2 \phi\right)-f^{-1}\partial_t^2 \phi + \partial_i\partial^i \phi-m^2 r^{2}\phi=0\,.
\ee
The standard prescription for computing the retarded Green's function is to find the solution $\phi^{\rm in}$ of the wave equation for such bulk fields with the infalling boundary condition at the horizon \cite{Son:2002sd,Nunez:2003eq,Policastro:2001yc,Policastro:2002se}, which ensures causality. Then, the holographic dictionary gives the retarded two-point function as the ratio of the coefficients of the two powers of $r$ near the AdS boundary (the normalizable mode and the non-normalizable mode) \cite{Witten:1998qj,Son:2002sd} 
\be
\phi^{\rm in}\sim \langle \mathcal{O}\rangle r^{-\Delta} + \phi_0 \, r^{\Delta-d}\,,\quad G_R\sim \fft{\delta \langle \mathcal{O}\rangle }{\delta \phi_0}\,.\label{eq: holographic dic}
\ee
This prescription is well studied and enjoys a wide range of applications. Note that from the asymptotic structure of the metric, the solution of the wave equation must decay in powers $r^{-\Delta-n d}$, with $n$ an integer. This dictionary therefore immediately suggests that the operators in the OPE that control holographic thermal retarded correlators are multi-stress-tensor operators $T^n$ with $\Delta' = n d$. Recently, it was pointed out that this resulting thermal retarded correlator develops a complex time singularity at $t_c = \beta (1+i)/2$ in $d=4$, which in frequency space corresponds to ``instanton-like" nonperturbative corrections of order $e^{-\omega \beta/2}$ \cite{Afkhami-Jeddi:2025wra}, and can also be understood from the asymptotic resummation of OPE coefficients \cite{Ceplak:2024bja}.

\subsection{Bouncing singularity from WKB method with $p=0$}
\label{subsec: WKB bouncing}

In this subsection, we review the WKB proposal of \cite{Afkhami-Jeddi:2025wra} which directly compute the nonperturbative, exponentially suppressed corrections in frequency space. We will fill in various technical details, in an attempt to keep the presentation self-contained.  

The wave equation of scalar field in momentum space $\phi=\int d\omega d^{d-1}p \phi_{\omega,p}(r)e^{-i\omega t+ip\cdot x}$ is
\be
\partial_r^2 \phi_{\omega,p}+ \left(\fft{5}{r}+\fft{f'}{f}\right)\partial_r \phi_{\omega,p}+ \fft{(\omega^2-p^2-m^2 r^2 f)}{r^4 f^2}\phi_{\omega,p}=0\,.\label{eq: bulk scalar eq p}
\ee
Here we will work only at $p=0$ (see \cite{Jia:2025jbi} for an analysis with nonzero $p$). This equation can be transformed into the Heun equation \cite{hortaccsu2018heun}, which then allows the retarded Green's function to be written in terms of Nekrasov-Shatashvili (NS) functions \cite{Aminov:2020yma,Bonelli:2022ten,Bonelli:2021uvf,Bianchi:2021mft,Bianchi:2021xpr,Fioravanti:2021dce,Consoli:2022eey,Dodelson:2022yvn,Aminov:2023jve,Bautista:2023sdf,Jia:2024zes} (see Appendix \ref{app: Heun} for a brief summary). Nevertheless, for planar black holes, analytic formulas are difficult to obtain in closed form using this approach, especially in the $\omega\rightarrow\infty$ limit, which is our main focus.

\begin{figure}[h]
\centering \hspace{0mm}\def\svgwidth{120mm}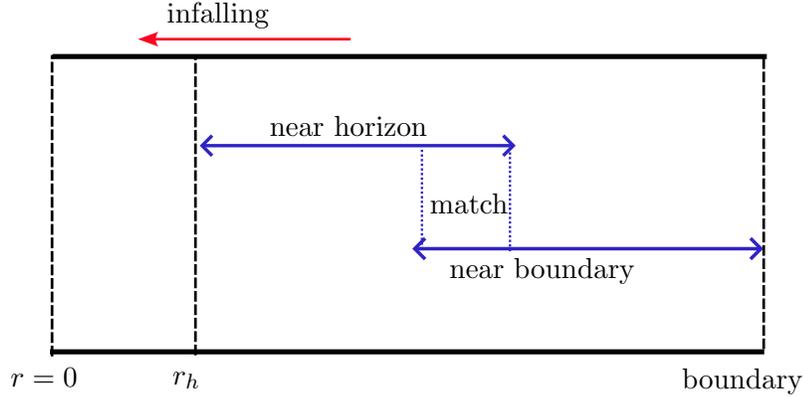
\caption{A schematic figure to illustrate the regime of validity of near-horizon solution with infalling boundary condition and near-boundary solution. The overlap region is enclosed by dashed blue line, where the matching procedure is performed.}
\label{fig: matching}
\end{figure}

A more practical approach is to solve the wave equation \eqref{eq: bulk scalar eq p} both asymptotically and near the horizon with infalling boundary conditions, and to match the solutions in an intermediate region so as to determine the retarded Green's function approximately at large scale separations, such as in an expansion in $1/\omega$ in our case. This matching method applies to both asymptotically AdS spacetimes \cite{Gubser:1997yh,Gubser:1997se,Klebanov:1999xv,Son:2002sd,Policastro:2002tn,Kovtun:2003wp,Kovtun:2004de,Teaney:2006nc,Blake:2019otz} and asymptotically flat spacetimes \cite{Mano:1996gn,Mano:1996mf,Mano:1996vt,Sasaki:2003xr,Ivanov:2022qqt,Saketh:2023bul,Saketh:2024juq}, and we illustrate this idea in Fig. \ref{fig: matching}. 

In the $1/\omega$ expansion, this approach is equivalent to the asymptotic OPE method \cite{Fitzpatrick:2019zqz,Fitzpatrick:2019efk} that will be discussed in the next section, as it systematically computes the $1/(\beta\omega)$ expansion perturbatively in the bulk and naturally extracts the retarded OPE (see the subsection \ref{subsec: OPE omega} for details of this approach). Nevertheless, this solution gives the retarded correlator only as an asymptotic series in $1/\omega$, and therefore misses nonperturbative corrections unless an appropriate resummation is performed.

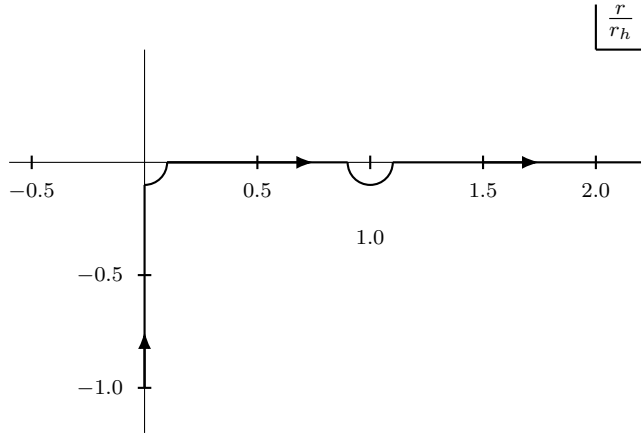
\begin{figure}[h]
\centering
\begin{tikzpicture}[scale=3, >=Latex]

    % --- Axis Lines (Thin Gray) ---
    % Extending slightly beyond the main ticks
    \draw[] (-0.6, 0) -- (2.2, 0);
    \draw[] (0, -1.2) -- (0, 0.5);
    % --- Ticks and Labels ---
    % X-axis
    \foreach \x in { -0.5, 0.5, 1.5, 2.0} {
        \draw[thick] (\x, -0.03) -- (\x, 0.03);
        \node[below, font=\scriptsize] at (\x, -0.05) {$\x$};
    }
    % Special handling for x=1.0 to avoid the arc
    \draw[thick] (1.0, 0.03) -- (1.0, -0.03);
    \node[below, font=\scriptsize, yshift=-5pt] at (1.0, -0.2) {1.0};

    % Y-axis
    \foreach \y in {-0.5, -1.0} {
        \draw[thick] (0.03, \y) -- (-0.03, \y);
        \node[left, font=\scriptsize] at (-0.05, \y) {$\y$};
    }
    % --- Main Paths (Thick Black with Arrows) ---
    % Define a style for putting an arrow in the middle of a path
    % 1. Horizontal Path: Origin -> Dip -> Right
    % Segment A: 0 to 0.8
    \draw[thick, black, ->] (0.1,0) -- (0.75, 0);
    \draw[thick,black] (0.1,0) -- (0.9,0);
    \draw[thick,black] (1.1,0) -- (2.2,0);
    
    % Segment B: The Dip (Semi-circle below x=1)
    % Center is (1,0), starts at 180 deg (0.8,0), goes counter-clockwise to 360 deg (1.2,0)
    \draw[thick, black] (0.9, 0) arc (180:360:0.1);
    
    % Segment C: 1.2 to 2.2
    \draw[thick, black,->] (1.5, 0) -- (1.75, 0);
    \draw[thick, black] (0,-1) -- (0,-0.1);

    % 2. Vertical Path: Origin -> Down
    % We stop at -1.3 to keep it neat
    \draw[thick, black, ->] (0,-1) -- (0, -0.75);
    % --- Angle Marker ---
    % Arc from negative y-axis (-90) to approx -45 degrees
    \draw[thick, black] (0, -0.1) arc (-90:0:0.1);
    % (Optional: add a tiny node or dot to show what it was measuring, 
    % but we keep it clean as requested)

    % --- Scale Legend (Top Right) ---
    \begin{scope}[shift={(2, 0.5)}]
        % Vertical bar
        \draw[thick, black] (0, 0) -- (0, 0.2); 
        % Horizontal base
        \draw[thick, black] (0, 0) -- (0.2, 0);
        % Label
        \node[anchor=south west, font=\normalsize] at (0, 0) {$\frac{r}{r_h}$};
    \end{scope}
\end{tikzpicture}
\caption{We choose the steepest descent contour that begins at $r=-ir_h$ and analytically continues to the real axis, reaching $r\to\infty$ after crossing the horizon below $r=r_h-i0$.}
\label{fig: contour}
\end{figure}

\begin{figure}[h]
\centering
\begin{tikzpicture}[>=Stealth]

    % --- Axes ---
    % Horizontal axis
    \draw[->, thick] (-4, 0) -- (8, 0);
    % Vertical axis
    \draw[thick] (0, -3) -- (0, 4);

    % --- Red Zig-Zag Cuts ---
    
    % 1. Central Vertical Cut (on positive y-axis)
    \draw[thick, red, decorate, decoration={zigzag, segment length=5, amplitude=2}] (0, 0) -- (0, 4);
    
    % 2. Top Left Cut (isolated)
    \draw[thick, red, decorate, decoration={zigzag, segment length=5, amplitude=2}] (-3, 2) -- (-3, 4);
    
    % 3. Top Right Cut (isolated)
    \draw[thick, red, decorate, decoration={zigzag, segment length=5, amplitude=2}] (3, 2) -- (3, 4);
    
     \draw[thick, red, decorate, decoration={zigzag, segment length=5, amplitude=2}] (5, 2) -- (5, 4);
     
      \draw[thick, red, decorate, decoration={zigzag, segment length=5, amplitude=2}] (6, 2) -- (6, 4);
      
      \draw[thick, red, decorate, decoration={zigzag, segment length=5, amplitude=2}] (5, -2) -- (5, -3);
     
      \draw[thick, red, decorate, decoration={zigzag, segment length=5, amplitude=2}] (6, -2) -- (6, -3);
    
    % 4. Bottom Left Cut (isolated)
    \draw[thick,red, decorate, decoration={zigzag, segment length=5, amplitude=2}] (-3, -2) -- (-3, -3);

    % 5. Bottom Right Cut (isolated)
    \draw[thick, red, decorate, decoration={zigzag, segment length=5, amplitude=2}] (3, -2) -- (3, -3);

    % --- Red Straight Line (on positive x-axis) ---
    \draw[thick, ->, blue] (0.2, 0) -- (8, 0);
    \node at (4,0.4) {Fourier transform};
    \node at (0.8,2) {Borel};

    % --- Red "X" Marks at Branch Points ---
    % Origin
    \node[red, font=\bfseries] at (0, 0) {$\times$};
    % Top Left bottom
    \node[red, font=\bfseries] at (-3, 2) {$\times$};
    % Top Right bottom
    \node[red, font=\bfseries] at (3, 2) {$\times$};
    % Bottom Left top
    \node[red, font=\bfseries] at (-3, -2) {$\times$};
    % Bottom Right top
    \node[red, font=\bfseries] at (3, -2) {$\times$};
    
    \node[red, font=\bfseries] at (5, 2) {$\times$};
    
     \node[red, font=\bfseries] at (6, 2) {$\times$};
     
      \node[red, font=\bfseries] at (5, -2) {$\times$};
             
        \node[red, font=\bfseries] at (6, -2) {$\times$};

    % --- Blue Contours (Integration Paths) ---

    % 1. The Central Contour (along the y-axis cut)
    % Vertical line going up on the right side of the y-axis
    \draw[thick,purple, ->] (0.2, 0) -- (0.2, 4);
    % Small arc connecting the real axis region to the vertical line
    \draw[thick, blue] (0.2, 0) arc (0:-90:0.2);
    % 2. The Top Right Contour (wrapping around the cut)
    % Downward line on the left side of the cut
    \draw[thick, purple] (2.8, 4) -- (2.8, 2);
    % Semi-circle loop at the bottom
    \draw[thick, purple] (2.8, 2) arc (180:360:0.2);
     \draw[thick, purple] (4.8, 2) arc (180:360:0.2);
      \draw[thick, purple] (5.8, 2) arc (180:360:0.2);
    % Upward line on the right side of the cut with arrow
    \draw[thick,purple, ->] (3.2, 2) -- (3.2, 4);
     \draw[thick, purple] (4.8, 4) -- (4.8, 2);
      \draw[thick, purple,->] (5.2, 2) -- (5.2, 4);
        \draw[thick, purple] (5.8, 4) -- (5.8, 2);
      \draw[thick, purple,->] (6.2, 2) -- (6.2, 4);

    % --- Legend "t" in the corner ---
    % Drawing the little corner marker manually
    \draw[thick] (4.5+3, 3.5) -- (4.5+3, 3.0) -- (5.0+3, 3.0);
    \node at (4.7+3, 3.25) {\Large $t$};
\end{tikzpicture}
\caption{The real time contour (blue) defining the Fourier transform of the retarded correlator $G_R(t)$ can be deformed into the purple contour: the segment along the imaginary axis implements the canonical Borel resummation, while the portions encircling the branch cuts generate the nonperturbative corrections.}
\label{eq: time contour}
\end{figure}
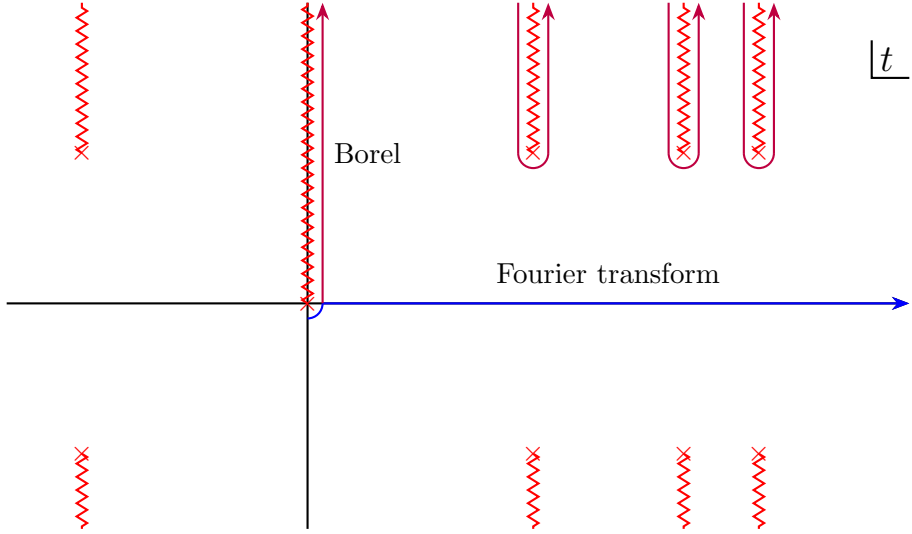

Recently, \cite{Afkhami-Jeddi:2025wra} proposed a WKB-based approach that enables the direct computation of nonperturbative corrections of the form $e^{-\#\omega}$. In order to understand the logic of the proposal, let us first recall that the retarded correlator in frequency space is given by the Fourier transform
\begin{equation}
G_R(\omega,p=0) = \int_0^{\infty} dt e^{i\omega t} G_R(t,p=0)\,.
\end{equation}
%In general, at large frequencies we expect $G_R(\omega,p=0)=G_R^{\rm pert}(\omega,p=0)+G_R^{\rm np}(\omega,p=0)$, where $G_R^{\rm pert}(\omega,p=0)$ is an asymptotic series in inverse powers of $\omega$, and $G_R^{\rm np}(\omega,p=0)$ contains the nonperturbative terms. Note that 
One may rotate the integration contour from the real axis to the imaginary axis, at the price of picking up the contributions from the possible singularities in the complex plane  (e.g., the contributions of contours encircling the branch cuts in the upper right quadrant, as shown in Fig \ref{eq: time contour}), which give nonperturbative terms at large $\omega$. Therefore   
\begin{equation}
\int_0^{\infty} dt e^{i\omega t} G_R(t,p=0) = \int_0^{\infty} d\tau e^{-\omega \tau} iG_R(i\tau,p=0)+{\rm nonperturpative~terms}
\end{equation}
On the other hand, one can see that the integral along the imaginary axis\footnote{To be more precise, the integration contour runs parallel to the imaginary axis, with a small positive shift in the real direction, as shown in Fig. \ref{eq: time contour}.} on the right-hand side above coincides with the canonical Borel resummation of the asymptotic series \eqref{eq: GR OPE}
\be
%[\mathcal{B}G_{R}(\omega,p=0)](\tau)=G_R(i \tau,p=0)\,,\quad 
\mathcal{S}_{\rm can}[G_R](\omega,p=0)=\int_0^\infty d\tau e^{-\omega \tau} (\mathcal{B}G_{R})(\tau)=\int_0^{\infty} d\tau e^{-\omega \tau} iG_R(i\tau,p=0)\,.
\ee
where $(\mathcal{B}G_{R})(\tau)$ denotes the Borel transform.\footnote{Recall that, for an asymptotic series of the form $f(x)=\sum_{k=0}^{\infty} b_k x^{k+a}$, the Borel transform is $(\mathcal{B}f)(\tau) = \sum_{k=0}^{\infty} \frac{b_k}{\Gamma\left(k+a\right)}\tau^{k+a-1}$, and the canonical Borel resummation of the asymptotic series is $\mathcal{S}_{\rm can}[f](x) =\int_0^{\infty} d\tau e^{-\tau/x} (\mathcal{B}f)(\tau)$. In the present case, the asymptotic series is $G_R^{\rm pert}(\omega,p=0)=\sum_{n=0}^{\infty} \hat{a}_n^{R\omega} (-i\omega)^{2\Delta-d-n d}$, and its Borel transform $(\mathcal{B}G_{R})(\tau) = i \sum_{n=0}^{\infty} \frac{\hat{a}_n^{R\omega}}{\Gamma\left(n d-2\Delta+d\right)} (i\tau)^{n d-2\Delta+d-1}=i G_R(i\tau,p=0)$, see \eqref{eq: GR t OPE} and \eqref{eq: GR OPE}.} Since it is given by a Laplace-type integral along a contour that contains no singularity, the large $\omega$ expansion of $\mathcal{S}_{\rm can}[G_R](\omega,p=0)$ only includes the perturbative series in powers of $1/\omega$, and therefore coincides with the perturbative part of $G_R(\omega,0)$
\be
G_R(\omega,0)=G_R^{\rm pert}(\omega,p=0)+G_R^{\rm np}(\omega,p=0)=\mathcal{S}_{\rm can}[G_R](\omega,p=0)+ G_R^{\rm np}(\omega,p=0)
%\simeq G_R^{\rm pert}(\omega,p=0)+\mathcal{O}(e^{-\# \beta \omega})\,.
\ee
The proposal of \cite{Afkhami-Jeddi:2025wra} is that the canonical Borel resummation $\mathcal{S}_{\rm can}[G_R](\omega,p=0)$ is computed in the bulk by solving the wave equation along a steepest descent contour that runs from the complex horizon $r=-ir_h$, where the solution of the wave equation is regular, crosses the horizon from below at $r_h - i0$, and extends to real $r \to \infty$ (see Fig.~\ref{fig: contour}). 
%in $r$ corresponds to the canonical Borel resummation of \eqref{eq: GR OPE} in thermal CFTs. 
%The contour runs .
%\be
%[\mathcal{B}G_{R}(\omega,p=0)](\tau)=G_R(i \tau,p=0)\,,\quad 
%\mathcal{S}_{\rm can}[G_R](\omega,p=0)=\int_0^\infty d\tau e^{-\omega \tau} %\mathcal{B}G_{R}(\tau)\,.
%\ee
%This is essentially the contour of the Fourier transform of $G_R(t,p=0)$ that is only along the positive imaginary axis, without wrapping around other branch cuts in the complex plane, while the physical retarded Green's function in frequency space is given by a full contour that receives possible nonperturbative contributions from branch cuts (see Fig \ref{eq: time contour}), giving 
%\be
%G_R(\omega,0)=\mathcal{S}_{\rm can}[G_R](\omega,p=0)+ G_R^{\rm np}(\omega,p=0)\simeq G_R^{\rm pert}(\omega,0)+\mathcal{O}(e^{-\# \beta \omega})\,.
%\ee

\begin{figure}[h]
\centering \hspace{0mm}\def\svgwidth{150mm}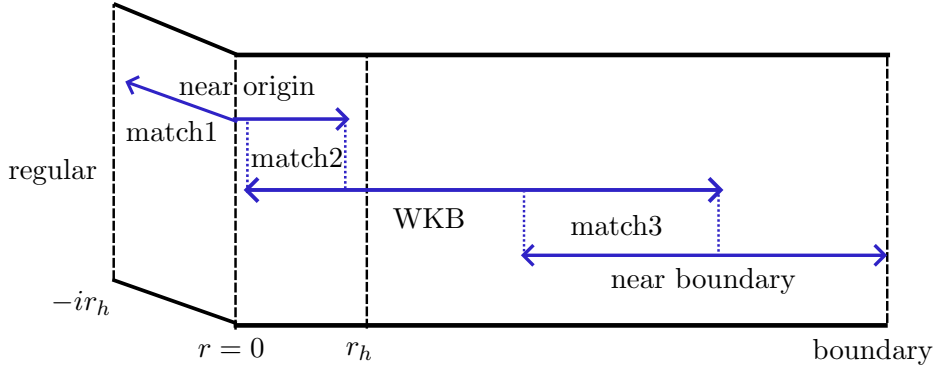
\caption{A schematic figure illustrating the procedure for constructing the steepest descent contour solution that is regular at $r=-i r_h$. A sequence of matching steps is performed to impose the regularity condition, connect to the WKB phase, and ultimately match to the near-boundary solution.}
\label{fig: matching np}
\end{figure}

In practice, at large $\omega$, obtaining such steepest descent solution that yields $\mathcal{S}_{\rm can}[G_R]$ requires a dedicated matching procedure. One proceeds by first analyzing the solution near the origin with fixed $x\sim\omega r^3/\mu$, where the usual WKB approximation breaks down, and imposing regularity as $x\to i\infty$ (equivalently, $\omega\to\infty$ at $r=-i$). This solution is then continued to real infinity $r\to\infty$ by following WKB geodesics. We illustrate this procedure in Fig.\ \ref{fig: matching np}. 

Since the WKB approximation used in the matching misses an unknown nonperturbative piece at large frequency, restoring this contribution allows one to write, at large $r$:
\be
\phi^{\rm steep}_\omega(r)=\phi^{{\rm in},{\rm pert}}_\omega(r)+\phi^{{\rm in},{\rm np}}_\omega(r)- A e^{-\# \beta \omega } \phi^{{\rm out}}(r)\,,\label{eq: steep structure}
\ee
where $\phi^{{\rm out}}(r)$ is an outgoing component near the boundary. 
By requiring that $\phi^{\rm steep}_{\omega}(r)$ corresponds to the canonical Borel resummation along the imaginary time contour, which implies that its large $\omega$ expansion should be purely perturbative (i.e. requiring $\phi^{\rm steep}_\omega(r)=\phi^{{\rm in},{\rm pert}}_\omega(r)$), we conclude that the physical infalling solution has a nonperturbatively small outgoing component near the boundary given by 
\be
\phi^{{\rm in},{\rm np}}_\omega(r) =  A e^{-\# \beta \omega } \phi^{{\rm out}}\,.\label{eq: np structure}
\ee
This then allows \cite{Afkhami-Jeddi:2025wra} to directly compute the nonperturbative corrections in the physical retarded correlator $G_R(\omega,p=0)$.

In the next subsections, we present the technical details of this construction.  

\subsubsection{Near-origin analysis}
\label{near-origin-sec}

Let's first analyze the near origin solution. In the large $\omega$ limit, keeping $x=\omega r^3/3$ fixed,\footnote{This scaling limit is suggested by analyzing the regime of validity of the large frequency  WKB expansion, which breaks down when $r\sim \omega^{-1/3}$, see eq. (\ref{WKB-sols}) below, where $r_s\sim -r^3/3$ for small $r$.}  the equation \eqref{eq: bulk scalar eq p} reduces to
\be
\partial_x^2 \phi_\omega + \fft{\partial_x \phi_\omega}{x}+\phi_\omega=\fft{12 x \partial_x \phi_\omega-(18x^2+m^2)\phi_\omega}{(3x)^{\fft{2}{3}}\omega^{\fft{4}{3}}}+\mathcal{O}\left(\omega^{-\fft{8}{3}}\right)\,.
\ee
We note that $x$ is related to the Tortoise coordinates at small $r$ by $x=-\omega r_s$, where
\be
r_s= \int_0^r \fft{dr'}{r'^2 f(r')}\,.
\ee
At leading order, the solution that is regular at $y=i^3 x \rightarrow\infty$ is
\be
\phi^{(0)}_\omega(y)= \fft{2i}{\pi} K_0(y)\,,
\ee
where the normalization is chosen just for convenience. To perform the analytic continuation and find the solution at $x\rightarrow\infty$, it is convenient to consider $y\rightarrow 0$, where the leading order solution gives $\sim -\log y+\log 2-\gamma_E$. The analytic continuation around $y\sim 0$ is straightforward $y\rightarrow e^{3 i \pi/2} x$, giving
\be
\lim_{x\rightarrow 0}\phi^{(0)}_\omega(x)=\fft{2i}{\pi} \left(-\log x + \log 2 -\gamma_E-\fft{3 i \pi}{2}\right)\,.
\ee
Evolving such initial data to large $x$ straightforwardly yields
\be
\phi^{(0)}_\omega(x)= H_0^{(1)}(x)-R^{(0)}H_0^{(2)}(x)\,,\quad R^{(0)}=-2\,,
\ee
where $R$ is referred to be as the reflection coefficient by crossing the singularity from complex horizon to horizon. This then provides approximate solution near the horizon $x\sim \omega /3 \rightarrow\infty$.

We can now go to the next-to-leading order in $1/\omega$
\be
\partial_x^2 \phi_\omega^{(1)} + \fft{\partial_x \phi_\omega^{(1)}}{x}+\phi_\omega^{(1)}=\fft{12 x \partial_x \phi_\omega^{(0)}-(18x^2+m^2)\phi_\omega^{(0)}}{(3x)^{\fft{2}{3}}\omega^{\fft{4}{3}}}:=J[\phi_\omega^{(0)}]\,.\label{eq: NL eq}
\ee
It is important to note that the next-to-leading-order correction scales as a power of $1/\omega^{4/3}$. Although the equation \eqref{eq: NL eq} is in general difficult to solve exactly, we can nevertheless extract the correct asymptotic behavior in the limit $x \to \infty$. The strategy is to first analyze the solution in the neighborhood of $y \to 0$, and then use this solution as initial data to evolve toward the large $x$ region. In real $y$, we have the ``Euclidean'' Green's function
\be
G_{\phi}(y,y')=y'\left(K_0(y')I_0(y)  \theta(y'-y)+K_0(y)I_0(y')\theta(y'-y)\right)\,.
\ee
We thus find
\be
& \phi^{(1)}(y)=-\left(\int_0^y dy' y' K_0(y')I_0(y) J[\phi^{(0)}(y')]+\int_y^\infty dy' y' K_0(y)I_0(y')J[\phi^{(0)}(y')]\right)\,,\nn\\
& \lim_{y\rightarrow 0}\phi^{(1)}(y)=\int_0^\infty dy' K_0(y')J[\phi(y')]=\fft{27+m^2}{7 \omega^{4/3}}C_0\,,\quad C_0=\fft{2^{\fft{2}{3}}\pi^2 \Gamma\left(\fft{2}{3}\right)}{3^{\fft{1}{6}}\Gamma\left(\fft{1}{6}\right)^2}\,.
\ee
The initial data at $x=0$ thus reads
\be
\lim_{x\rightarrow 0}\phi_\omega(x)=\lim_{x\rightarrow 0}\phi_\omega^{(0)}(x)+\fft{2i}{\pi}\fft{27+m^2}{7 \omega^{4/3}}C_0\,.\label{eq: initial cond}
\ee
We thus then solve \eqref{eq: NL eq} at $x\rightarrow\infty$ with initial condition given by \eqref{eq: initial cond} and $\phi^{(0)}=H_0^{(1)}+2 H_0^{(2)}$. Generically, the solution takes the form of
\be
\phi^{(1)}(x)=a_0(x) H^{(1)}_0(x)+b_0(x) H^{(2)}_0(x)\,,\quad b_0'= -\fft{H_0^{(1)}a_0'}{H_0^{(2)}}\,.
\ee
The problem then boils down to solving the differential equations for $(a_0,b_0)$
\be
a_0'= \fft{J[\phi^{(0)}(x)] H_0^{(2)}(x)}{W[x]}\,,\quad W[x]=H_0^{(1)}\fft{d}{dx}H_0^{(2)}-H_0^{(2)}\fft{d}{dx}H_0^{(1)}=-\fft{4i}{\pi x}\,,
\ee
which are solved by
\be
a_0=a_{0c}+ \int_0^x dx' \fft{J[\phi^{(0)}(x')]H_0^{(2)}(x')}{W[x']}\,,\quad b=b_{0c}- \int_0^x dx' \fft{J[\phi^{(0)}(x')]H_0^{(1)}(x')}{W[x']}\,.\label{eq: ab0 sol}
\ee
Taking $x\rightarrow 0$, we can match with the initial data \eqref{eq: initial cond} to determine $(a_{0c},b_{0c})$
\be
a_{0c}=b_{0c}=i\fft{(24+7m^2)}{7\pi}\,.
\ee
We can then take the limit $x \to \infty$ in \eqref{eq: ab0 sol} to extract the large $x$ behavior of the solution. The integrals give hypergeometric functions and Meijer G functions. To detect their large $x$ asymptotic behavior, we use their integral representations and deform the contour to pick up appropriate residues. We defer the details of this computation to Appendix \ref{app: integrals}, and simply quote the final results here. The resulting terms grow polynomially at large $x$, and this growth is in fact essential for matching onto the WKB phase, as will be demonstrated in the next subsection. We find
\be
& \lim_{x\rightarrow \infty}a_0=-\fft{\sqrt{3}(24+7m^2)C_0}{7\pi} +i\left(\fft{1}{7}(3x)^{\fft{7}{3}}+ \fft{15+4m^2}{8}(3x)^{\fft{1}{3}}\right)\,,\nn\\
& \lim_{x\rightarrow \infty}b_0=-\fft{\left(\sqrt{3}-9i\right)(24+7m^2)C_0}{7\pi} -i\left(\fft{2}{7}(3x)^{\fft{7}{3}}+ \fft{15+4m^2}{4}(3x)^{\fft{1}{3}}\right)\,.
\ee
Therefore, we have up to $1/\omega^{3/4}$ order
\be
\lim_{x\rightarrow\infty}\phi_\omega(x)\sim \phi_\omega\left(\fft{r^3}{3 r_h^3}\omega \right)\Big|_{r\sim r_h}\sim -\left(1-\fft{\sqrt{3}(24+7m^2)C_0}{7\pi \omega^{\fft{4}{3}}} \right)\left(u_1(x) -R(\omega) u_2(x)\right)\,,\label{eq: sol horizon}
\ee
where the independent two solutions are
\be
u_1=\left(1+\fft{i}{\omega^{\fft{4}{3}}}\left(\fft{1}{7}(3x)^{\fft{7}{3}}+ \fft{15+4m^2}{8}(3x)^{\fft{1}{3}}\right)\right)H_0^{(1)}\,,\quad u_2=u_1^\ast\,.
\ee
The reflection coefficient reads
\be
R(\omega)=R^{(0)}+ \omega^{-\fft{4}{3}}R^{(1)}\,,\quad R^{(1)}=-\fft{3\sqrt{3}(24+7m^2)C_0}{14\pi}e^{i\fft{\pi}{3}}\,,\label{eq: bulk scalar R}
\ee
which agrees with \cite{Afkhami-Jeddi:2025wra}.

\subsubsection{WKB connections and the retarded correlator}
\label{subsec: WKB connection}

With the steepest descent contour solution near the horizon \eqref{eq: sol horizon}, we can now follow the WKB geodesic to evolve it to infinity and match asymptotic solutions.

It is straightforward to solve the WKB solution for \eqref{eq: bulk scalar eq p} with $p=0$. The WKB ansatz is
\be
\phi_{\omega}\sim e^{\omega S^{(-1)}+S^{(0)}+\fft{1}{\omega}S^{(1)}+\cdots}\,.
\ee
The equation \eqref{eq: bulk scalar eq p} becomes
\be
& \omega^2\left(\left(\partial_r S^{(-1)}\right)^2+\fft{1}{r^4 f^2}\right)+\omega\left(\partial_r^2 S^{(-1)}+\fft{5 f+r f'}{r f} \partial_r S^{(-1)}+2 \partial_r S^{(-1)}\partial_r S^{(0)}\right)\nn\\
& +\left(\partial_r^2 S^{(0)}+\fft{5 f+r f'}{r f} \partial_r S^{(0)}+\left(\partial_r S^{(0)}\right)^2+2\partial_r S^{(-1)}\partial_r S^{(1)}-\fft{m^2}{r^2 f}\right)+\cdots=0\,.
\ee
The leading order gives the time shift\footnote{For a null geodesic, one has $dt=\pm dr/(r^2 f(r))$.}
\be
S^{(-1)}=\pm i\int_0^r dr' \fft{1}{r'^2 f(r')}=\pm i r_s\,,\label{eq: geodesic}
\ee
Going to the next-to-leading order, we find the two WKB solutions
\be
\phi_\omega^{\pm {\rm WKB}}(r)\sim r^{-3/2} \exp\left(\pm i\omega r_s \mp i \fft{(15+4m^2)r^4-3}{8\omega r^3}+\cdots \right)\,.
\label{WKB-sols}
\ee
To establish the connections with \eqref{eq: sol horizon}, it is necessary to track the WKB geodesic all the way to falling into the horizon by taking $r/r_h\rightarrow 0$ but still $r/r_h\gg \omega^{-1/3}$, giving
\be
\lim_{\omega^{-1/3}\ll r/r_h\ll 1}\phi_\omega^{\pm {\rm WKB}}\sim e^{\mp i x}\left(1\mp \fft{i}{\omega^{\fft{4}{3}}}\left(\fft{1}{7}(3x)^{\fft{7}{3}}+ \fft{15+4m^2}{8}(3x)^{\fft{1}{3}}\right)+\cdots\right)\,.
\ee
Matching to \eqref{eq: sol horizon} yields
\be
\phi_{\omega}^{\rm steep}\sim \phi_\omega^{- {\rm WKB}}+R(\omega)\phi_\omega^{+ {\rm WKB}}\,.\label{eq: steep in WKB}
\ee
We now simply track the WKB to asymptotic infinity at $r\rightarrow\infty$ by noting
\be
\lim_{r\rightarrow\infty}\phi_\omega^{\pm {\rm WKB}}(r)\sim \exp\left(\pm i \fft{\beta}{4}(1+i) \omega\mp i \fft{\omega}{r}\right)\,,\label{eq: WKB 1/r}
\ee
where we perform the integral in \eqref{eq: geodesic} along the contour passing through the horizon slightly from below in the complex plane. %Therefore, it appears that the time shift that bounces back from horizon (i.e., the bouncing singularity) actually comes from the WKB solution asymptotically and shows up in retarded correlators by the matching. 
On the other hand, the near boundary solution is 
\be
\lim_{r\rightarrow\infty}\phi_\omega^{\rm steep}\left(z=\fft{\omega}{r}\right)\sim \fft{1}{r^2}\left(\mathcal{C}_1 H_\nu^{(1)}(z)+\mathcal{C}_2 H_\nu^{(2)}(z)\right)\,,\label{eq: steepes H12}
\ee
where $\Delta=\nu+d/2$. The near boundary solution is required to match to \eqref{eq: steep in WKB} with \eqref{eq: WKB 1/r} in the limit of $z\rightarrow \infty$, where
\be
\lim_{z\rightarrow\infty}H_\nu^{(1)}(z)\sim \fft{(1-i)}{\sqrt{\pi z}}e^{iz -\fft{i \pi \nu}{2}}\,,\quad \lim_{z\rightarrow\infty}H_\nu^{(2)}(z)\sim \fft{(1+i)}{\sqrt{\pi z}}e^{-iz +\fft{i \pi \nu}{2}}\,.
\ee
We find
\be
\mathcal{C}_2\sim - e^{-i\pi\nu -\fft{\beta}{2}\left(1-i\right)\omega} R(\omega) \mathcal{C}_1\,.\label{eq: sol}
\ee
We note that $H_\nu^{(1)}$ is the perturbative infalling solution $\phi_\omega^{{\rm in},{\rm pert}}$, therefore, as we promised in \eqref{eq: steep structure} and \eqref{eq: np structure}, \eqref{eq: steepes H12} and \eqref{eq: sol} combine to give
\be
\lim_{r\rightarrow\infty}\phi_\omega^{\rm in}\sim \fft{1}{r^2}\left(H_\nu^{(1)}\left(\fft{\omega}{r}\right)+e^{-i\pi\nu-\fft{\beta}{2}(1-i)\omega}R(\omega)H_\nu^{(2)}\left(\fft{\omega}{r}\right)\right)\,.\label{eq: phi in infall}
\ee
Applying the holographic dictionary \eqref{eq: holographic dic} to \eqref{eq: phi in infall}, we obtain the following retarded correlator
\be
G_R(\omega,p=0)&=-\fft{\omega^{2\nu}}{2\sin \pi \nu} \fft{R(\omega)e^{i \pi \nu}- e^{\fft{\beta}{2}(1-i)\omega}}{R(\omega)-e^{i\pi \nu+ \fft{\beta}{2}(1-i)\omega}}\nn\\
&=\omega^{2\nu}\left(-\fft{e^{-i\pi \nu}}{2\sin \pi \nu}+i e^{-i\pi \nu -\fft{\beta}{2}(1-i)\omega}R(\omega)+\mathcal{O}(e^{-\beta \omega})\right)\,,\label{eq: GR final}
\ee
where we normalize the retarded correlator by
\be
G_W^{>,{\rm vac}}=\omega^{2\nu}\,,\quad G_W^{>,{\rm vac}}=2 \fft{{\rm Im}\,G_R^{\rm vac}}{1-e^{-\beta \omega}}\,.\label{eq: normalization}
\ee
In \eqref{eq: GR final}, we neglect higher order power corrections $1/\omega$ and higher non-perturbative corrections $e^{-\beta \omega}$, and the reflection coefficient $R$ can be found in \eqref{eq: bulk scalar R}. 

From the first line in \eqref{eq: GR final}, we can deduce the asymptotic quasinormal modes
\be
\omega_q=\omega_{0q}+\fft{3\sqrt{3}(7\nu^2-4)e^{\fft{7 i\pi}{12}}C_0}{7\sqrt{2}\pi^2\omega_{0q}^{\fft{4}{3}}}\,,\quad \omega_{0q}=\fft{\pi}{\beta}\left((i-1)+2(1-i)n+(1-i)\nu+\fft{(1+i)\log 2}{\pi}\right)\,,
\ee
where the leading asymptotic quasinormal modes $\omega_{0q}$ agree with \cite{Natario:2004jd,Dodelson:2023vrw}. 

The second line of \eqref{eq: GR final} agrees with \cite{Afkhami-Jeddi:2025wra} and indeed presents instanton-like nonperturbative behavior $\sim e^{-\beta\omega/2}$. Note that, formally, the exponential term can be written as $e^{i\omega \Delta t}$, where $\Delta t=2\int_0^{\infty} dr/r^2f(r)=\beta(1+i)/2$ is twice the complex time shift of a null geodesic from the boundary to $r=0$. However, we emphasize that in the detailed calculation we performed above, the nonperturbative tail arises from the near-boundary WKB phase through the matching with the interior solution derived in section \ref{near-origin-sec}, rather than from assuming the validity of the simple WKB geodesic picture over the full region from the boundary to the singularity. 
%, and thus counting the time shift from the boundary $r\sim\infty$ to the black hole singularity $r=0$.

\subsubsection{Bouncing singularity and OPEs}
\label{subsec: OPE-from-sing}

We can now Fourier transform the second line of \eqref{eq: GR final} to the time domain and identify the singularity in complex time, dubbed the bouncing singularity. The nonperturbative tail immediately shows that this singularity is located at $t_c=\beta/2(1+i)$. We find
\be
G_R(t,p=0)\sim -\fft{e^{-2i\pi \nu}}{\pi}\left(\fft{\Gamma\left(2\nu+1\right)}{(t-t_c)^{2\nu+1}}-\left(\fft{\pi}{\beta}\right)^{\fft{4}{3}} \fft{3\sqrt{3}(7\nu^2-4)C_0}{14\pi} \fft{\Gamma\left(2\nu-\fft{1}{3}\right)}{(t-t_c)^{2\nu-\fft{1}{3}}}\right)+ {\rm regular}\,.\label{eq: GR in t}
\ee
Note that such a singularity lies beyond the conventional analyticity strip of Wightman functions and therefore resides on the second sheet\footnote{The conventional analyticity strip of Wightman function at finite temperature is $-\beta < {\rm Im} t< 0 $.}, but there is no known principle that forbids such singularities and branch cuts for the retarded two-point functions. In fact, the analytic structure of the retarded Green function in frequency space remains elusive beyond real positive time.\footnote{See \cite{Correia:2025enx} for an attempt with asymptotically flat black holes.} Therefore, this bouncing singularity in the retarded Green function should really be understood as a signal of the nonperturbative tail $e^{-\beta \omega/2}$ in frequency space.

In addition, we note that \eqref{eq: GR in t} exhibits singularities in power law rather than other complicated singular functions. This provides a strong prediction for the OPE data from multi-stress-tensor exchange $T^n$ asymptotically at large $n$, as expected from the Tauberian theorem \cite{Qiao:2017xif,Marchetto:2023xap}. To show this, we note that
\be
\sum_{n=1} x^{-1-2\nu} n^{\alpha-m}x^{n d}=x^{-1-2\nu} {\rm Li}_{m-\alpha}(x^d)\,,\quad x=\fft{t}{t_c}\,,
\ee
whose form is suggested by the retarded OPE in the time domain \eqref{eq: GR t OPE}. The task is to find appropriate coefficients for different positive integers $m$ such that, when summed, they produce a single pure power singularity of the form $1/(t - t_c)^\#$. Matching to \eqref{eq: GR in t}, we find
\be
\lim_{n\rightarrow\infty}\hat{a}_{4n}^{Rt}=\fft{4^{n+1+2\nu}}{\pi} e^{-in\pi} n^{2\nu}\left( r_{2\nu,\nu}+ \fft{3\sqrt{3}C_0\left(\fft{\pi}{2}\right)^{\fft{1}{3}}(7\nu^2-4) r_{2\nu-\fft{4}{3},\nu}}{112 n^{\fft{4}{3}}}+\mathcal{O}(n^{-\fft{8}{3}})\right)\,,\label{eq: a4n asymp}
\ee
where
\be
r_{\alpha,\nu}& =1
+ \frac{\alpha \bigl(\alpha - 1 - 4 \nu\bigr)}{2d n }
+ \frac{\alpha (\alpha - 1) \bigl(3 \alpha^2 - \alpha (24 \nu + 7) + 48 \nu^2 + 24 \nu + 2\bigr)}{24d^2 n^2 }
\nn\\
& + \frac{(-2 + \alpha) (-1 + \alpha) \alpha (-1 + \alpha - 4 \nu) \bigl(\alpha^2 + 8 \nu (1 + 2 \nu) - \alpha (3 + 8 \nu)\bigr)}{48d^3 n^3}+\mathcal{O}(n^{-4})\,.
\label{r-alpha-nu}
\ee
This also reproduces the results in \cite{Afkhami-Jeddi:2025wra}.

%\subsection{Bouncing singularity from asymptotic OPE method with $p=0$}
\subsection{Asymptotic OPE method with $p=0$}
\label{sec: bouncing from OPE p0}

\subsubsection{OPE in time domain}
\label{subsec: OPE t}

In this subsection, we turn to another approach that was originally proposed in \cite{Fitzpatrick:2019zqz,Fitzpatrick:2019efk} to study the holographic OPE from near-boundary expansions in position space with Euclidean signature. See also \cite{Li:2019tpf} for applications to various theories. We refer to this as the asymptotic OPE method, which has also been used to probe bouncing singularities at $x=0$ \cite{Ceplak:2024bja}. In this subsection, we slightly modify this method to apply to momentum space at $p=0$, with analytic continuation used to compute the retarded OPE.

We consider the Euclidean black hole, obtained from \eqref{eq: bh metric} by the analytic continuation $t \rightarrow \pm i \tau$. We then study the equation of motion for a scalar field in momentum space and focus on the zero modes by setting $p = 0$
\be
r^{d-1} \left(\left((d+1)f+r f'\right)\partial_r \phi_{p=0}+r \partial_r^2 \phi_{p=0}\right)+f^{-1}\partial_\tau^2 \phi_{p=0} -m^2 r^{d-2}\phi_{p=0}=0\,.
\ee
It is more convenient to work in the coordinates $(r,w)$, where $w^2 = 1 + r^2 \tau^2$. Our strategy is to first determine the bulk-to-boundary propagator from this equation and then take the limit $r \rightarrow \infty$ to extract the Euclidean two-point functions. To this end, we begin by recalling that in pure AdS, corresponding to $\mu = 0$, the bulk-to-boundary propagator is known.
\be
G_{b\partial}^{\rm AdS}(\tau,p=0)\sim %\fft{\Gamma(\Delta)}{\pi^{\fft{d}{2}}\Gamma\left(\Delta-\fft{d}{2}\right)}
 \int d^dx \fft{r^{\Delta}}{(1+r^2(\tau^2+x^2))^{\Delta}}\sim \fft{r^{\Delta-3}}{(1+r^2 \tau^2)^{\Delta-\fft{3}{2}}}\,.\label{eq: Gbpartial}
\ee
We thus then define
\be
\phi_{p=0}(r,w)=\fft{r^{\Delta-d+1}}{w^{2\Delta-d+1}}\Psi(r,w)\,,
\ee
where, by the structure of the equation, we expect
\be
\Psi(r,w)=\sum_{n=0} \fft{\mu^n g_n(w)}{r^{n d}}\,,\label{eq: expansion varphi}
\ee
which encodes the perturbations around pure AdS with small but fixed $\mu$. Therefore, the logic here is different from that in the previous sections. We are genuinely solving the bulk scalar equation of motion around pure AdS perturbatively by performing the Fefferman-Graham expansion in $1/r$, which does not rely on any knowledge of the existence of a horizon. In this approach, there is no sharp distinction between computations performed in a black hole background, which yield thermal-two-point functions, and those performed in a compact star background, which yield heavy-heavy-light-light four-point functions. We will interpret our results as thermal two-point function and leave more comments to section \ref{sec: universality}. 

Via the holographic dictionary, the structure of the expansion in \eqref{eq: expansion varphi} immediately suggests that multi-stress-tensor operators $T^n$ are exchanged in the $\mathcal{O} \times \mathcal{O}$ OPE. We can also clearly see the physical interpretation of the coordinate $w$: the bulk fields are expanded near the boundary $r \to \infty$ in a way that probes the OPE limit $\tau \to 0$ while keeping $w$ fixed. It turns out that in even dimensions such as $d = 4$ \cite{Fitzpatrick:2019zqz,Fitzpatrick:2019efk,Li:2019tpf}, the functions $g_n(w)$ can be solved in terms of truncated polynomials\footnote{This is not true in odd dimensions, where a finite truncation requires special values of $\Delta$.}
\be
g_{n}(w)=\sum_{k=-2n}^{d n} Q_{nk}w^k\,.
\ee
We can now see that taking $w^2=1+\tau^2 r^2$ and $r\rightarrow\infty$ computes the Euclidean two-point function with correct OPE structure \eqref{eq: G_E p0}, where
\be
\hat{a}_{dn}=Q_{n,dn}\pi^{d n}\,,
\ee
where the normalization of two-point function is determined by the ``identity operator OPE'' $Q_{00}$. To agree with the convention in previous sections, we consider
\be
Q_{00}=\fft{\Gamma\left(2\nu+1\right)}{2\pi}\,.
\ee
It is important to note that, in general, $\mathcal{O} \times \mathcal{O}$ OPEs contain double-trace operators, such as $\mathcal{O} (\epsilon \cdot \partial)^J \partial^{2n} \mathcal{O}$, which are not captured by this method \cite{Fitzpatrick:2019zqz,Fitzpatrick:2019efk,Li:2019tpf}. Nevertheless, at leading order in $1/N$, Fourier transforming to momentum space $p$ eliminates all double-trace operators \cite{Manenti:2019wxs}, even without analytically continuing to the retarded correlator. This can be seen in \eqref{eq: a def} by noting that the Gamma function factors have zeros at the double-trace dimensions $\Delta' = 2\Delta_{\mathcal{O}} + J + 2n$. The underlying reason is nicely explained in \cite{Manenti:2019wxs}, where it is shown that the spatial Fourier transform naturally picks out the spatial discontinuity, yielding again the commutator $[\mathcal{O}, \mathcal{O}]$. Similarly, as we will explain shortly, analytically continuing to the retarded correlator in position space also eliminates all double-trace operators due to the appearance of factors of $\sin\left((\Delta' - 2\Delta)\pi/2\right)$, as shown in section \ref{subsec: thermal correlator}. The OPE coefficients in \eqref{eq: GR t OPE} are thus given by
\be
\hat{a}_{dn}^{Rt}=2Q_{n,dn}\pi^{d n} \cos\left(\fft{d+dn-2\Delta}{2}\pi\right) \,.
\ee

Practically, we can build a recursion equation for $Q_{kn}$ and solve it iteratively easily to around $n\sim 650$ in few minutes. The low-lying OPE coefficients in $d=4$ read:
\be
\hat{a}_{4}^{Rt}&=\frac{\cos(\pi \Delta)\Gamma(-3 + 2 \Delta) (-4 + \Delta) (-1 + \Delta) \Delta \pi^4}{10 \pi (-7 + 2 \Delta) (-5 + 2 \Delta)}\,,\nn\\
 \hat{a}_{8}^{Rt}&=\frac{\cos(\pi \Delta)\Gamma(-3 + 2 \Delta) (-1 + \Delta) \Delta (96 + 80 \Delta + 240 \Delta^2 - 95 \Delta^3 + 9 \Delta^4)\pi^8}{1800 \pi (-11 + 2 \Delta) (-9 + 2 \Delta) (-7 + 2 \Delta) (-5 + 2 \Delta)}\,,\nn\\
 \hat{a}_{12}^{Rt}&=\frac{\cos(\pi \Delta)\Gamma(-3 + 2 \Delta) (-8 + \Delta) (-1 + \Delta) \Delta}{1638000 \pi (-15 + 2 \Delta) (-13 + 2 \Delta) (-11 + 2 \Delta) (-9 + 2 \Delta) (-7 + 2 \Delta) (-5 + 2 \Delta)} \nn\\
 & \times \left(17280 + 40992 \Delta + 38512 \Delta^2 + 15260 \Delta^3 + 5075 \Delta^4 - 2912 \Delta^5 + 273 \Delta^6\right)\pi^{12}\,.\label{eq: a Rt}
\ee
We emphasize that in momentum space with $p=0$, we find no singularity in integer $\Delta$ that arises from mixing into double-trace operators, which is different from the OPE coefficients in position space. The underlying reason can be traced to \eqref{eq: a def}, where the Gamma function prefactors have zeros for all double-trace dimensions and develop spurious poles in half-integer $\Delta$.

\begin{figure}[t]
    \centering
    \includegraphics[width=0.8\textwidth]{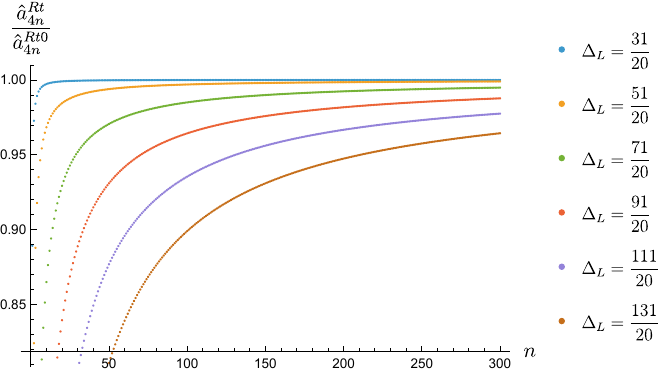}
    \caption{The convergence of $\hat{a}_{4n}^{Rt}/\hat{a}_{4n}^{Rt0}$ as $n$ increases, where $\hat{a}_{4n}^{Rt}$ is obtained numerically. Comparisons are shown for several values of $\Delta$. }
    \label{fig: fit bulk OPE p0}
\end{figure}

We then follow the methodology of \cite{Ceplak:2024bja}: compute $a_{4n}^{Rt}$ to sufficiently high $n$ and fit the asymptotic tails. At leading order, we find
\be
\hat{a}_{4n}^{Rt0}\sim \fft{4^{n+1+2\nu}}{\pi}e^{-i n\pi}n^{2\nu}\,,
\ee
which indeed agrees with the leading order result in \eqref{eq: a4n asymp}. We refer to Fig \ref{fig: fit bulk OPE p0} for illustration of this comparison and asymptotic convergence by plotting $\hat{a}_4^{Rt}/\hat{a}_{4n}^{Rt0}$.

To figure out subleading corrections, we computed $\hat{a}_{4n}^{Rt}$ to be around $n=650$ and then fit in polynomials
\be
{\rm fit}[n]= \fft{\hat{a}_{4n}^{Rt}}{\hat{a}_{4n}^{Rt0}}-1\,.
\ee
Presumably, this fitting procedure is not stable if we assume that all subleading corrections scale as $1/n^{\rm integer}$. Nevertheless, from the WKB analysis in the previous section, we know that a term scaling as $1/n^{4/3}$ must be present. Taking this into account, we perform the fit using polynomials below
\be
{\rm fit}[n]=\sum_{ij}^{i+\fft{4}{3}j=N_{\rm max}} \fft{c_{ij}}{n^{i+\fft{4}{3}j}}\,.
\ee
With this choice, the fit becomes stable and not only reproduces \eqref{eq: a4n asymp}, but also predicts the numerical coefficient of the $1/n^{8/3}$ correction. For example, we obtain\footnote{We choose $\Delta=31/20,51/20,71/20\cdots$ as our examples by following \cite{Ceplak:2024bja}.}
\be
 \Delta=\fft{31}{20}:\quad {\rm fit}[n]&=\frac{0.01125}{n} - \frac{0.079408}{n^{4/3}} - \frac{0.000311719}{n^2} - \frac{0.0317742}{n^{7/3}} - \frac{0.000408248}{n^{8/3}} - \frac{0.0000145283}{n^3} \nn\\ 
& - \frac{0.0110452}{n^{10/3}} - \frac{0.000503563}{n^{11/3}} + \frac{0.0521248}{n^4} - \frac{0.00372373}{n^{13/3}}+\cdots\,,\nn\\
 \Delta=\fft{51}{20}:\quad {\rm fit}[n]&=- \frac{0.28875}{n} - \frac{0.0578841}{n^{4/3}} + \frac{0.00318828}{n^2} - \frac{0.00579645}{n^{7/3}}  - \frac{0.0127745}{n^{8/3}} + \frac{0.000156331}{n^3}\nn\\
& - \frac{0.00150079}{n^{10/3}} - \frac{0.0119246}{n^{11/3}} - \frac{0.0528955}{n^4} - \frac{0.000458247}{n^{13/3}}+\cdots\,.\label{eq: fits bulk scalar some}
\ee
We can also perform the fit for all other values of $\Delta$, which we present in Appendix~\ref{app: more fits p=0}. This extends the result of \cite{Afkhami-Jeddi:2025wra} for $\Delta=4$ to other $\Delta$, which we also reproduce below
\be
 \Delta=4:\quad {\rm fit}[n]&=- \frac{2.5}{n} + \frac{0.737964}{n^{4/3}} + \frac{2.1875}{n^2} - \frac{1.55792}{n^{7/3}}  + \frac{0.361533}{n^{8/3}} - \frac{0.78125}{n^3} \nn\\
& + \frac{0.99648}{n^{10/3}}  - \frac{0.461959}{n^{11/3}} + \frac{0.270864}{n^4} - \frac{0.164287}{n^{13/3}}+\cdots\,.
\ee
We can easily verify that the numerical values match \eqref{eq: a4n asymp} for the relevant powers. In addition, we predict numerical coefficients for the $n^{-8/3}$ and $n^{-11/3}$ terms for several values of $\Delta$. Therefore, we can easily reproduce \eqref{eq: GR in t} that demonstrates the bouncing singularity in complex time at $t_c=\beta/2(1+i)$.

\subsubsection{OPE in frequency domain}
\label{subsec: OPE omega}

Now we show that we can also directly compute the retarded OPE in frequency space $\hat{a}_{4n}^{R\omega}$ by following the recursive approach of \cite{Afkhami-Jeddi:2025wra}. This method is the Lorentzian, frequency space counterpart of the Euclidean asymptotic OPE method used previously.

The strategy is similar, we aim to solve \eqref{eq: bulk scalar eq p} by taking $r\rightarrow\infty$, $\sqrt{\omega^2-p^2}\rightarrow\infty$ with fixed $z=\sqrt{\omega^2-p^2}/r$, where in Lorentzian signature we impose the perturbative infalling boundary condition at ``horizon'' $z\rightarrow\infty$. We still take $p=0$ for simplicity. At leading order, we are simply considering the pure AdS, and the solution is
\be
\phi_{\omega}^{\rm AdS}\sim z^2 H_\nu^{(1)}(z)\,.
\ee
As we mentioned in the subsection \ref{subsec: WKB connection}, this solution obeys the perturbative infalling boundary condition at the horizon, where in this regime we have $z\sim \omega/r_h\to\infty$. However, we emphasize that the situation here is different from the nonperturbative WKB proposal discussed previously, where interior data behind the horizon are explicitly imported \cite{Afkhami-Jeddi:2025wra}. In our large $\omega$ approximation, the limit $z\to\infty$ itself does not distinguish between the horizon at $r=r_h$ and the origin at $r=0$. Indeed, the solution $H_\nu^{(1)}$ is the conventional bulk-to-boundary propagator in pure AdS without any horizon, which becomes $K_\nu$ in Euclidean signature that is regular at the Poincare horizon $r\to 0$, and reproduces \eqref{eq: Gbpartial} after Fourier transform.

We thus take the same viewpoint as in the asymptotic OPE method in time domane, where we perform perturbation theory around pure AdS in $1/r$ with fixed $\omega/r$. The ambiguous interpretation of the boundary condition, namely whether it is imposed at the horizon that is not manifest in the computation, at the origin, or at any finite $r$, precisely reflects the long distance effective field theory logic: long distance physics (the asymptotic analysis we perform) and short distance physics (structures at the horizon and in the interior) are factorized. We will return to this point in section \ref{sec: universality}.

The general lesson from \cite{Afkhami-Jeddi:2025wra} is that, similar to the subsection \ref{subsec: OPE t}, the solution has the structure
\be
\phi_{\omega}\sim z^2 H_\nu^{(1)}(z)+  \sum_{n=1}\fft{\mu^n}{\omega^{4n}} \left( g_{1n}(z)H_{\nu}^{(1)}(z)+g_{2n}(z)H_{\nu+1}^{(1)}(z)\right)\,,\label{eq: phi expand frequency}
\ee
where in even dimensions $g_{in}$ are rational polynomials
\be
g_{1n}=\sum_{m=2}^{2(n+1)}g_{1nm}z^{2m}\,,\quad g_{2n}=\sum_{m=1}^{2(n+1)} g_{2nm}z^{2m-1}\,.
\ee
For example, we find at order $\mu$
\be
 g_{11}= \frac{1}{10} z^4 (-1 + \nu) (-4 + \nu^2)+\frac{1}{10} z^6 (1 + 2 \nu) \,,\quad g_{21}= \frac{1}{5} z^3 (-4 + 5 \nu^2 - \nu^4)+ \frac{1}{10} z^5 (4 - \nu^2) -\frac{z^7}{5} \,.
\ee
We can systematically do this to sufficiently high $n$, and then use the dictionary \eqref{eq: holographic dic} to find the retarded two-point function. Up to the order $\mu^2$ with the conventional normalization we take in \eqref{eq: normalization}, we find
\be
& G_R(\omega,p=0)=-\fft{e^{-i \pi \nu}}{\sin \pi \nu}\omega^{2\Delta-4}\Bigg(1 + \frac{2 \Delta (\Delta-1)(\Delta-2)(\Delta-3)(\Delta-4)}{5 \omega^4} \nn\\
& + \frac{2 \Delta (\Delta-1)(\Delta-2)(\Delta-3)(\Delta-4)(\Delta-5) \left(9 \Delta^4 - 95 \Delta^3 + 240 \Delta^2 + 80 \Delta + 96\right)}{225 \omega^8}+\cdots \Bigg)\,.
\ee
We can thus extract $\hat{a}_{4n}^{R\omega}$ easily and convert $\hat{a}_{4n}^{Rt}$ using \eqref{eq: a def} and \eqref{eq: GR OPE}. Noting the identity such as
\be
\fft{\Gamma(-2\nu) \Gamma(2\nu+1)\cos(\Delta \pi)}{\pi}=-\fft{1}{2\sin (\pi \nu)}\,,
\ee
we find agreements with \eqref{eq: a Rt} and also for all larger values of $n$.

We can also see how nonperturbative tail $e^{-\beta/2 \omega}$ arises in this case. The factor $\Gamma(d+\Delta-2\Delta)$ for $\Delta=n d$ further exponentiates at large $n$, then we need to perform the integral roughly as
\be
\int dn \fft{4^{n+1+2\nu}}{\pi}e^{-i n \pi} n^{2\nu}\frac{e^{(4n - 2\nu) (\log(4n - 2\nu) - 1)} \sqrt{\pi}}{\sqrt{2n}} (-i \omega)^{2\nu-4n}\,.
\ee
We find the saddle point of $n$ is
\be
n^*=\fft{1}{8}\left((1-i)\omega+1- \fft{1+i}{4\omega}+\cdots \right)\,.
\ee
Evaluating on the saddle-point of this integral yields precisely the phase $e^{-\beta/2(1-i)\omega}$.

\subsection{Comments on asymptotic OPE with $x=0$}
%case from asymptotic OPE}
\label{subsec: x0 comments}

It is also interesting to study the retarded correlator in position space, particularly $G_R(t,x=0)$. In this case, only the asymptotic short time OPE method is available \cite{Fitzpatrick:2019zqz,Fitzpatrick:2019efk,Li:2019tpf} to detect the bouncing singularity, as demonstrated in \cite{Ceplak:2024bja,Ceplak:2025dds}. However, the analysis in \cite{Ceplak:2024bja} was restricted to the leading large $n$ asymptotic behavior of the multi-stress-tensor OPE coefficients. By contrast, the discussions in \cite{Afkhami-Jeddi:2025wra} and in subsection \ref{subsec: OPE t} indicate that a stable fit capturing higher order $1/n$ corrections is in fact possible. For example, as is clear from \eqref{eq: OPE x0} and \eqref{eq: G_E p0}, passing from momentum space to position space does not modify the power law contributions associated with each multi-stress-tensor operator. We therefore continue to expect the presence of $1/n^{4/3}$ corrections.

We refer the details of the asymptotic OPE method at $x=0$ to Appendix \ref{app: OPE x=0}. Here we briefly review the key results of \cite{Ceplak:2024bja} and explain how their tail formula can be improved. The conjectured asymptotic formula from \cite{Ceplak:2024bja} is (which we refer to as CLPV)
\be
\lim_{n\rightarrow\infty}a_{4n}^{\rm CLPV}=\frac{\pi \Delta^2 (\Delta - 1) 4^{n + 2\Delta - 1} n^{2\Delta - 3} e^{-i n \pi} \csc(\pi \Delta)}{5 \Gamma\left(2\Delta + \frac{3}{2}\right)}\,.\label{eq: a CLPV}
\ee
We first emphasize that the analytic continuation to retarded correlator gives (see section \ref{subsec: thermal correlator})
\be
a_{4n}^{R,{\rm CLPV}}=-2\sin\left(\pi \Delta\right)a_{4n}^{{\rm CLPV}}\,.
\ee
We then again try to perform the fit for several $\Delta$ using the following format
\be
{\rm fit}^{x=0}[n]= \fft{a_{4n}^{R}}{a_{4n}^{R,{\rm CLPV}}}-1\,,\quad {\rm fit}^{x=0}[n]=\sum_{ij}^{i+\fft{4}{3}j=N_{\rm max}} \fft{c_{ij}^{x=0}}{n^{i+\fft{4}{3}j}}\,.
\ee
The numerical evaluation in the $x=0$ case is much more demanding than in the $p=0$ case, because there are two indices that must be evolved in the recursion, as explained in Appendix \ref{app: OPE x=0}. We therefore only reach a maximal value of $n_{\rm max}=200$. However, and more importantly, we find that the discrepancy from ${\rm fit}^{x=0}[n]\to 0$ pointed out in \cite{Ceplak:2024bja} persists and stabilizes even at the largest accessible $n$, and is neither a numerical artifact nor a $1/n$ correction. Instead, it reflects the fact that the leading asymptotic ansatz \eqref{eq: a CLPV} itself is incomplete, in the sense that its normalization misses a nontrivial dependence on $\Delta$. For example, we find
\be
& \Delta=\fft{31}{20}: \quad {\rm fit}^{x=0}=0.0160119 - \frac{0.0647708}{n} + \frac{0.146813}{n^{4/3}} - \frac{0.0183803}{n^2}+\cdots\,,\nn\\
& \Delta=\fft{51}{20}: \quad {\rm fit}^{x=0}=0.00624633 - \frac{1.87539}{n} - \frac{0.700078}{n^{4/3}} + \frac{0.896431}{n^2}+\cdots\,,\nn\\
& \Delta=\fft{71}{20}: \quad {\rm fit}^{x=0}=0.00125602 - \frac{4.66961}{n} - \frac{1.90805}{n^{4/3}} + \frac{8.06402}{n^2}+\cdots\,.\label{eq: fits x=0 original}
\ee
We thus need to include the constant piece into the leading behavior of $a_{4n}^{R}$
\be
a_{4n}^{R0}=a_{4n}^{R,{\rm CLPV}}\times (1+ F(\nu))\,, \quad {\rm fit}^{x=0,{\rm new}}[n]=\fft{a_{4n}^{R}}{a_{4n}^{R0}}-1\,.
\ee
By performing the calibration for several $\Delta$, we find
\be
F(\nu)\sim -0.0140082 + \frac{0.963958 + 1.1603 \nu + 0.464078 \nu^2 + 0.061585 \nu^3}{\left(\frac{5}{2} + \nu\right)^4}\,.\label{eq: F fit}
\ee
Furthermore, we have for example
\be
& \Delta=\fft{31}{20}: \quad {\rm fit}^{x=0,{\rm new}}=- \frac{0.06375}{n} + \frac{0.144499}{n^{4/3}} - \frac{0.0180907}{n^2}+\cdots\,,\nn\\
& \Delta=\fft{51}{20}: \quad {\rm fit}^{x=0,{\rm new}}=- \frac{1.86375}{n} - \frac{0.695732}{n^{4/3}} + \frac{0.890867}{n^2}+\cdots\,,\nn\\
& \Delta=\fft{71}{20}: \quad {\rm fit}^{x=0,{\rm new}}=- \frac{4.66375}{n} - \frac{1.90565}{n^{4/3}} + \frac{8.0539}{n^2}+\cdots\,.\label{eq: fits =0}
\ee
We can now verify that our fits for the $1/n$ and $1/n^2$ terms very accurately agree with
\be
{\rm fit}^{x=0,{\rm new}}=r_{2\nu+1,\nu+\fft{3}{2}}+\fft{\tilde{F}(\nu)}{n^{\fft{4}{3}}}r_{2\nu-\fft{1}{3},\nu+\fft{3}{2}}+\mathcal{O}(n^{-8/3})\,.\label{eq: fit new formula}
\ee
We also predict the coefficients of the $1/n^{3/4}$ behavior for several values of $\Delta$, and rough fits give
\be
\tilde{F}(\nu)\sim \frac{\pi^{7/3} \Gamma\left(\frac{2}{3}\right)}{32 \cdot 6^{2/3} \Gamma\left(\frac{1}{6}\right) \Gamma\left(\frac{13}{6}\right)} \left(-6.10944 - 26.7248 \nu - 6.01147 \nu^2\right)\,.\label{eq: Ft fit}
\ee
Note that the fits in \eqref{eq: F fit} are performed by taking $\Delta = 31/20 + i$ with $i = 0,1,\cdots,5$, while the fits in \eqref{eq: Ft fit} are performed by taking $\Delta = 31/20, 51/20, 71/20$. We have also checked other values of $\Delta$ with the same order of magnitude, $\Delta \sim \mathcal{O}(1)$. It is possible that these formulas does not capture the correct behavior for larger values of $\Delta$, which we leave to future exploration. We record our fitted results for additional values of $\Delta$ in Appendix \ref{app: OPE x=0}.

\section{Thermal Wilson line correlators}
\label{sec: Wilson line}

In this section, we turn to the discussion of thermal correlation functions on Wilson lines using the holographic description in terms of fundamental strings in AdS.

\subsection{Wilson lines in thermal CFT}

Wilson loop are important observables in gauge theories. They are essential for capturing and classifying both propagating and confined degrees of freedom in a fully gauge invariant manner. In addition, they provide an effective description of the dynamics of heavy probe particles coupled to gauge fields (for example, heavy-quark effective theory \cite{Neubert:2005mu}).

The prototype we have in mind is to study Wilson lines in conformal gauge field theories, more specifically in $\mathcal N=4$ supersymmetric Yang-Mills theory in $d=4$ at strong coupling, using holography. Even though our calculations and results can be applied to more general line defects with a holographic dual, to be concrete we focus on the Wilson-Maldacena line, which preserves $16$ of the $32$ supercharges of the $\mathcal N=4$ superconformal group \cite{Maldacena:1998im}
\be
W=\fft{1}{N}{\rm Tr}\, P e^{\int_{\mathcal{C}} d\tau \left( i \dot{x}^\mu  A_\mu+ \Phi_a \theta^a |\dot{x}|\right)}\,,
\ee
where $\Phi_a$ are the six SYM scalars, and $\theta^a$ a unit vector. Let us first review the ``vacuum'' case and later generalize to finite temperature. It is convenient to choose $x^\mu$ to align with the Euclidean time direction, $x^0 = \tau$, and $\theta^a$ to align with the sixth direction, so that only $\Phi^6$ appears in the Wilson line. It is then clear that this line preserves an $\mathrm{SO}(5)$ subgroup of the $\mathrm{SO}(6)$ R symmetry, corresponding physically to rotations of the remaining decoupled scalars $\Phi^a$ with $a=1,\cdots,5$. It also preserves an $\mathrm{SO}(2,1)\times \mathrm{SO}(3)$ subgroup of the $\mathrm{SO}(2,4)$ conformal symmetry, where $\mathrm{SO}(3)$ describes rotations transverse to the line, and $\mathrm{SO}(2,1)$ is the residual one dimensional conformal symmetry along the line. Hence this system provides an example of a conformal line defect, and the result 1d defect CFT has been studied extensively in the literature \cite{Erickson:2000af,Drukker:2006xg,Cooke:2017qgm,Sakaguchi:2007ba,Giombi:2017cqn,Beccaria:2017rbe,Giombi:2018qox,Giombi:2018hsx,Beccaria:2019dws,Giombi:2020amn}. 

Our main object of interest are the correlation functions of local operators inserted on the line, which are organized by the one dimensional conformal symmetry of the defect CFT \cite{Drukker:2006xg,Sakaguchi:2007ba,Cooke:2017qgm}%although generically in a nonlocal manner 
\be
\langle\langle \mathcal{O}(\tau_1)\cdots \mathcal{O}(\tau_n)\rangle\rangle:=\langle {\rm Tr}\, P[ \mathcal{O}(x(\tau_1))\cdots \mathcal{O}(x(\tau_n))e^{\int iA+\Phi_6}]\,,
\ee
where we use the notation $\langle\langle \cdots \rangle\rangle$ to denote correlation functions of operators inserted on the Wilson line. The simplest multiplet is a short representation of $\mathrm{OSp}(4^\ast|4)$ with protected scaling dimensions: three displacement operators $D_i$ with $\Delta_D = 2$, and five scalar operators $\Phi_a$ with $\Delta_\Phi = 1$, where
\be
D_i =i F_{\tau i}+D_i \Phi_6\,.
\ee
%for example in QCD, where the $\Phi^6$ coupling is absent, but their dimensions are no longer protected. 
Physically, the correlation functions of displacement operators describe deformations of the straight line or circular contour, and hence describe how the trajectory of the heavy quark probe is deformed and deflected by the surrounding transverse radiation.

Our main goal is to study such a Wilson line and its 1D defect CFT correlators in $\mathcal N=4$ SYM at finite temperature, focusing on the retarded two-point function of the displacement operator. Physically, this describes a heavy quark moving in a quark-gluon plasma. In Euclidean signature, the Wilson line aligns with the thermal circle and therefore becomes a Polyakov loop \cite{Polyakov:1978vu}, where we have for the two-point function of an operator $\mathcal{O}$ inserted on the defect
\be
G_E^{\rm defect}(\tau)=\langle\langle \mathcal{O}(\tau)\mathcal{O}(0)\rangle\rangle_\beta :=\langle {\rm Tr}\, P_\beta \mathcal{O}(x(\tau))\mathcal{O}(x(0))e^{\int iA+\Phi_6}]\,.
\ee
To study real time dynamics, we perform an analytic continuation precisely as we described in section \ref{subsec: thermal correlator} but now for the 1D defect CFT, similarly for all OPE discussions which are now defect OPE along the Wilson line \cite{Giombi:2017cqn,Giombi:2018hsx,Beccaria:2019dws}. The geometric picture of this analytic continuation is that the Polyakov loop extends into two copies of the Wilson line, living on the forward timefold and the backward timefold of the Schwinger-Keldysh contour, with appropriate boundary conditions imposed by the trace \cite{Rajagopal:2025ukd,Rajagopal:2025rxr}. For example, formally we have
\be
G_R^{\rm defect}(t)=i \langle\langle [\mathcal{O}(t),\mathcal{O}(0)]\rangle\rangle_\beta \theta(t):=i \langle {\rm Tr}\, P_{\rm SK}[ [\mathcal{O}(x(t)),\mathcal{O}(x(0))] \theta(t) e^{\int iA+\Phi_6}]\,.
\ee
We will be mainly interested in the case where the operators $\mathcal{O}(t)$ are the displacement operators $D^i(t)$. The physical significance of thermal Wilson line retarded correlators of displacement operators is clear and intuitive: they precisely describe, in real time, how a heavy quark responds to transverse drag in a quark-gluon plasma \cite{Gubser:2006qh,Liu:2006he,Gubser:2006nz,Casalderrey-Solana:2011dxg}.

Even though for concreteness we have in mind the well-studied case of the half-BPS Wilson line, our holographic calculations below can also be applied to more general setups. For instance, one can consider the ordinary, non-supersymmetric Wilson loop operator without scalar couplings \cite{Alday:2007he, Polchinski:2011im, Beccaria:2017rbe}. More generally, one can consider any line defect in a holographic CFT, as long as it has an effective description in terms of a string worldsheet in AdS. The displacement operators, which as we review below are dual to transverse fluctuation of the string, can be defined for any line defect and have protected dimension $\Delta_D=2$ even in the absence of supersymmetry.\footnote{In the presence of a line defect, the stress tensor satisfies $\partial_{\mu}T^{\mu i} = \delta^{(d-1)}(x_{\perp}) D^i$, where $D^i$ are the displacement operators. Since the stress tensor has protected dimension equal to $d$, this fixes the scaling dimension of the displacement operator to be $\Delta_D=2$.}

\begin{comment}
In general, such displacement operators can be defined even in the absence of supersymmetry. For instance, for the ordinary Wilson loop without scalar coupling, they are just given by the field strength term above. For a conformal line defect, their dimension remains protected even in the absence of supersymmetry, as this is a consequence of the broken translational invariance.\footnote{In the presence of a line defect, the stress tensor satisfies $\partial_{\mu}T^{\mu i} = \delta^{(d-1)}(x_{\perp}) D^i$, where $D^i$ are the displacement operators. Since the stress tensor has protected dimension equal to $d$, this fixes the scaling dimension of the displacement operator to be $\Delta_D=2$.}

More generally, we can replace the Wilson line by other defects and study correlation functions of operators inserted on the defects \cite{Zarembo:2002an,Drukker:2007qr,Beccaria:2017rbe,Beccaria:2018ocq,Barrat:2024aoa}. Our results for displacement operator capture the universal and leading contribution for thermal retarded two-point function of displacement operators on all defects.
\end{comment}   

\subsection{Static string and asymptotic AdS$_2$}

We now set up the holographic computation of the retarded two-point functions on a thermal Wilson line at strong coupling using string theory in AdS$_5 \times S^5$. In the AdS/CFT dictionary, Wilson lines correspond to open string minimal surfaces in the bulk ending on the contour that defines the line on the boundary. For the $1/2$ BPS Wilson line in the vacuum, the minimal surface is AdS$_2$ embedded in AdS$_5$, localized at a point on $S^5$. As expected, the one dimensional defect CFT is therefore dual to AdS$_2$ on the string worldsheet \cite{Sakaguchi:2007ba,Gomis:2006sb,Drukker:2000ep,Faraggi:2011bb,Giombi:2017cqn}. The elementary bosonic fluctuations around the static string are described by three massive scalars with $m^2 = 2$ and five massless scalars with $m^2 = 0$, corresponding to the displacement operators and the SYM scalar fluctuations respectively.

For the thermal Wilson line, we consider string theory on the planar Schwarzschild black hole in AdS$_5$, which describe the $\mathcal N = 4$ SYM theory on $R^{1,3}$ at finite temperature
\be
ds^2=-r^2 f(r) dt^2 + \fft{dr^2}{r^2 f(r)}+r^2 dx^i dx^i + \fft{dy^a dy^a}{(1+y^2/4)^2}\,,\quad f(r)=1-\fft{\mu}{r^4}\,,
\ee
where $x^i$, $i=1,2,3$ are the spatial coordinates at the boundary, and $y^a$, $a=1,\ldots, 5$ the $S^5$ coordinates. 

The bosonic part of the superstring action in this background is given by the Nambu-Goto action
\begin{equation}
S_B = T \int d^2\sigma \sqrt{-{\rm det}_{\alpha\beta}\left[ 
-r^2 f(r) \partial_\alpha X^0 \partial_\beta X^0+\fft{\partial_\alpha X^r \partial_\beta X^r}{r^2 f(r)}  + r^2  \partial_\alpha X^i \partial_\beta X^i + \fft{\partial_\alpha Y^a \partial_\beta Y^a}{(1+Y^2/4)^2}
\right]
}
%(g_{\alpha\beta})}\\
\end{equation}
%where
%\be 
%g_{\alpha\beta} = -r^2 f(r) \partial_\alpha X^0 \partial_\beta X^0+\fft{\partial_\alpha X^r \partial_\beta X^r}{r^2 f(r)}  + r^2  \partial_\alpha X^i \partial_\beta X^i + \fft{\partial_\alpha Y^a \partial_\beta Y^a}{(1+Y^2/4)^2}\,
%\ee 
%\be
%S_B=\fft{1}{2}T \int d^2\sigma \sqrt{-h}h^{\alpha\beta}\left(\left(-r^2 f(r) \partial_\alpha X^0 \partial_\beta X^0+\fft{\partial_\alpha X^r \partial_\beta X^r}{r^2 f(r)}  + r^2  \partial_\alpha X^i \partial_\beta X^j \right)+ \fft{\partial_\alpha Y^a \partial_\beta Y^b}{(1+Y^2/4)^2}\right)\,,
%\ee
where the string tension is $T=\sqrt{\lambda}/(2\pi)$, and $\sigma^{\alpha}$ denotes the worldsheet coordinates.  The minimal surface which describes the Wilson line at the boundary is the static string extending in the $(t,r)$ directions. Explicitly, the classical solution is 
%We consider the minimal surface ending on the Wilson line $X^0=t$, fitting the plane $(t,r)$. More concretely we have
\be
X^0=t\,,\quad X^r=r\,,\quad X^i=Y^a=0\,.
\ee
The induced geometry of the worldsheet is given by the asymptotically AdS$_2$ metric
\be
ds^2_2=-r^2 f(r) dt^2 + \fft{dr^2}{r^2 f(r)}\,.
\label{AdS2-bh}
\ee
Note that the worldsheet has a horizon at $r=r_h$ inherited from the horizon of the bulk geometry. 

To develop perturbation theory on the worldsheet, one may fix a static gauge where the longitudinal coordinates $X^0$ and $X^r$ do not fluctuate. The transverse fluctuations away from $X^i = 0$ and $Y^a = 0$ then define massive and massless modes, corresponding to the displacement operators and the SYM five scalar operators $\Phi^a$, respectively. Expanding the Nambu-Goto action perturbatively in these fluctuations we can obtain an effective 2d theory in the AdS$_2$ ``black hole" (\ref{AdS2-bh}), which describes the 1D thermal defect CFT. In this paper, we restrict ourselves to the quadratic terms, leaving interactions to future work. To this order, the action for the fluctuations is
\be
S_B=\fft{1}{2} \int dt dr \sqrt{-h} \left(h^{\alpha\beta}\partial_\alpha \phi^i \partial_\beta \phi^i +\left(r f'(r)+2f(r)\right) \phi^i \phi^i + h^{\alpha\beta}\partial_\alpha \varphi^a \partial_\beta \varphi^a + \text{interactions} \right)\,,\label{eq: string action}
\ee
where $\phi^i = \delta X^i / r$ and $\varphi^a = \delta Y^a$, and $h_{\alpha\beta}$ is the induced worldsheet metric (\ref{AdS2-bh}). 

To compute the retarded correlator of displacement operators, we then follow the prescription reviewed in section \ref{sec: thermal correlator bulk} and study the wave equation for the massive scalar $\phi^i$ with infalling boundary condition at the horizon. From this point on, we will strip off the transverse index $i$, as the dependence of the two-point function on this index is trivial. The crucial difference compared to the bulk scalar field wave equation, in addition to working in the asymptotic AdS$_2$ space, is that the scalar fluctuations now have a coordinate dependent mass, as seen in (\ref{eq: string action}). As we will see below, this does not cause major difficulties, and both the WKB and asymptotic OPE analyses can be carried out in a similar way as what we described in the case of the bulk scalar field.  We leave comments on the massless scalar $\varphi^a$ to subsection \ref{subsec: massless}, where we show that in this case the retarded two-point function can be solved exactly.

\begin{comment}
We emphasize that even without having the $1/2$BPS Wilson line in SYM in mind as the UV theory, we can still consider our set up to study asymptotic AdS$_2$, which corresponds to computing 1D defect CFT for any possible defects, from bottom up perspective. For example, we can easily generalize our results to the case of AdS$_2$ embedded in AdS$_{d+1}$ and include more matter fields and higher derivatives interactions (for example, consider effective string theory in AdS \cite{Gabai:2025hwf,Gabai:2026myo}), which could be understood as a generic line defect in d-dimensions.
\end{comment}

%\section{Thermal Wilson line retarded correlator from holography}
\section{Retarded correlator on the Wilson line from holography}
\label{sec: thermal Wilson line correlators}

In this section, we discuss in detail the calculation of the thermal retarded correlator of displacement operators on the Wilson line. Our focus will be on the large frequency regime and the appearance of bouncing singularities, but for completeness we also briefly discuss the small frequency regime in the next subsection. 

% effectively computing 1D correlators in thermal defect CFT. 
From \eqref{eq: string action}, we find the following wave equation
\be
f(r) \left( m(r)^2 r^2 \phi - r^4 f'(r) \partial_r \phi \right) - r^3 f(r)^2 \left( 2 \partial_r \phi + r \partial_r^2 \phi \right) + \partial_t^2 \phi=0\,,
\ee
where we have $m(r)^2=r f'(r)+2f(r)$ for the massive modes (corresponding to displacement operators) and $m(r)\equiv 0$ for massless modes (corresponding to SYM scalars). In frequency space, we have
\be
\partial_r^2 \phi_\omega + \left( \frac{2}{r} + \frac{f'}{f} \right) \partial_r \phi_\omega +\left(\frac{\omega^2}{r^4 f^2} - \frac{m(r)^2}{r^2 f}\right)\phi_\omega  =0\,.\label{eq: eq string phi}
\ee
Despite the position dependent mass, we note that this equation can also be transformed to take the form of the Heun equation, see Appendix \ref{app: Heun}. 

\subsection{Small frequency and stochastic string}
\label{subsec: small frequency}

For completeness, let us first discuss the small frequency expansion of the retarded correlator. Physically, this is the stochastic regime in which a heavy quark behaves as a Brownian particle in the quark-gluon plasma: the intrinsic time scale of the plasma is much shorter than the response time scale of the heavy quark. In this limit, the thermal retarded correlator encodes the coefficients governing the near equilibrium response, including transport, polarizability and diffusion data. Holographically, transverse string fluctuations experience dissipation and thermal noise sourced by the worldsheet horizon, so the endpoint dynamics becomes stochastic \cite{Son:2009vu,Caron-Huot:2011vtx,Chakrabarty:2019aeu}.

We find the solution to \eqref{eq: eq string phi} at small $\omega$ with infalling boundary condition at the horizon has the structure
\be
\phi_\omega\sim \left(\fft{r-1}{r}\right)^{-i\fft{\omega}{4}} r\left(1+\sum_{n=1} \omega^n \psi_n(r)\right)\,.
\ee
Plugging this into the wave equation and solving for $\psi_n(r)$ up to $\omega^2$ order, we find
\be
& \psi_1=-\frac{i \pi}{8} - \frac{i}{4} \log 2 - \frac{1}{4} \log(1 - i r) + \frac{1}{4} \log(1 + i r) - \frac{i}{4} \log(r) + \frac{i}{4} \log(1 + r)\,,\nn\\
& \psi_2=\fft{1}{2}\psi_1^2+\frac{\pi}{8} + \frac{\log 2}{4} + \fft{1-i}{4} \log(1 - i r) +\fft{1+i}{4} \log(1 + i r) - \frac{1}{2} \log(1 + r)\,.
\ee
Using the holographic dictionary \eqref{eq: holographic dic} and the normalization \eqref{eq: normalization}, we find
\be
G_R^{\rm defect}(\omega)=\fft{i \omega}{2}-\fft{\omega^2}{2}+\mathcal{O}(\omega^3)\,.\label{eq: GR small}
\ee
This agrees with \cite{Son:2009vu} by changing the normalization condition.

\subsection{Defect bouncing singularity from WKB method}
\label{subsec: defect bouncing WKB}

We now turn to the large-$\omega$ limit and adapt the WKB analysis \cite{Afkhami-Jeddi:2025wra} presented in subsection \ref{subsec: WKB bouncing} to the thermal Wilson line case. Physically, this limit probes the short-time structure of the interaction between probe heavy quarks and the quark-gluon plasma, resolving microscopic constituents such as individual quarks and gluons rather than hydrodynamic modes. Consequently, these retarded correlators are expected to be dominated by vacuum physics \cite{Caron-Huot:2009ypo}. As in the case of bulk scalars, we will see that the retarded correlators in this regime display the nonperturbative exponentially suppressed terms that correspond to the bouncing singularities of $G_R(t)$ at complex time. 

Near the singularity $r=0$, the wave equation now reads
\be
\partial_x^2 \varphi_\omega+ \fft{\partial_x \varphi_\omega}{x} +\left(1- \fft{1}{36 x^2}\right)\varphi_\omega =-\frac{2 \left( (9 x^2 - 1) \varphi_\omega(x) - 6 x \partial_x \varphi_\omega(x) \right)}{3^{2/3} x^{2/3} \omega^{4/3}}\,,
\ee
where we defined $\varphi_\omega=\sqrt{x}\phi_\omega$. At leading order in $1/\omega$, we find the solution (regular at large $y$) 
\be
\varphi^{(0)}_\omega(y)=\frac{(1 + i)(i + \sqrt{3})}{\sqrt{2} \pi} K_{\fft{1}{6}}(y)\,.
\ee
The analytic continuation to the real axis, in a similar way as discussed in section \ref{subsec: massless}, then gives
\be
\varphi_\omega^{(0)}(x)=H_{1/6}^{(1)}(x) + \frac{1}{2} \left( 3 - i\sqrt{3} \right) H_{1/6}^{(2)}(x)\,.
\ee
Therefore, in this case we find the leading reflection coefficient
\be
R^{(0)}=-\frac{1}{2} \left( 3 - i\sqrt{3} \right)=-\sqrt{3}e^{-i\pi/6} \,.
\ee
Following the same method in section \ref{subsec: massless}, at the subleading order $\sim \omega^{-4/3}$, we find that the initial data is
\be
\lim_{x\rightarrow 0}\varphi_\omega(x)=\lim_{x\rightarrow 0}\varphi_\omega^{(0)}(x)+\frac{(1 + i)(i + \sqrt{3})}{\sqrt{2} \pi \omega^{\fft{4}{3}}}e^{\fft{i \pi}{4}}C_{\fft{1}{6}} x^{\fft{1}{6}}\,,\quad C_{\fft{1}{6}}=\frac{3^{1/3} \pi \Gamma\left(-\frac{7}{6}\right) \Gamma\left(\frac{2}{3}\right)}{2^{1/6} \Gamma\left(\frac{1}{6}\right)^2}\,.
\ee
Consequently, the solution is
\be
\lim_{x\rightarrow\infty}\varphi_\omega(x)\sim-\left(1 + \frac{4 \times 2^{1/6} C_{\fft{1}{6}}}{\omega^{4/3} \Gamma\left(-\frac{1}{6}\right)}\right) \left(u_1(x) - R u_2(x)\right)\,,
\label{interior-sol}
\ee
where
\be
u_1=\left(1+\fft{i}{\omega^{\fft{4}{3}}}\left(\fft{1}{7}(3x)^{\fft{7}{3}}+ (3x)^{\fft{1}{3}}\right)\right)H_{\fft{1}{6}}^{(1)}\,,\quad u_2=u_1^\ast\,.
\ee
From this, we read off the reflection coefficient at subleading order to be
\be
R(\omega)=R^{(0)}+\omega^{-\fft{4}{3}}R^{(1)}\,,\quad R^{(1)}=\frac{2i \times 2^{1/6} \left(-3i + \sqrt{3}\right)}{\Gamma\left(-\frac{1}{6}\right)}C_{\fft{1}{6}}\,.\label{eq: defect R reflection}
\ee

We can then move to establish the WKB connections from the interior to the near boundary region. In this case, the WKB solutions of the wave equation (\ref{eq: eq string phi}) are
\be
\phi_{\omega}^{\pm {\rm WKB}}\sim \exp\left(\pm i \omega r_s \mp \fft{i (3r^4-1)}{3\omega r^3}\right)\,.
\ee
Note that the leading order term in the exponent is given by the time shift, just as for bulk scalars. To match with the near-origin solution, we note that
\be
\lim_{\omega^{-1/3}\ll r/r_h\ll 1}\phi_\omega^{\pm {\rm WKB}}\sim e^{\mp i x}\left(1\mp \fft{i}{\omega^{\fft{4}{3}}}\left(\fft{1}{7}(3x)^{\fft{7}{3}}+(3x)^{\fft{1}{3}}\right)+\cdots\right)\,,
\ee
which can be matched with (\ref{interior-sol}). Finally, using as before the large $r$ limit of the WKB phase (\ref{eq: WKB 1/r}), and noting that the near-boundary solutions are linear combinations of $r^{-1/2}H_{\nu=3/2}^{(1)}(\omega/r)$ and $r^{-1/2}H_{\nu=3/2}^{(2)}(\omega/r)$, corresponding to the asymptotic value of the mass $m^2=2$,\footnote{The Hankel functions in this case reduce to elementary functions.} we find, following the same logic as described in section \ref{subsec: WKB connection}  
%Noting also that asymptotically we have the mass $m^2=2$ ($\nu=3/2$), we find
\be
\lim_{r\rightarrow\infty}\phi_\omega^{\rm in}\sim  \frac{(1-i)(i+\sqrt{3}) e^{\frac{i \omega}{r}} \sqrt{\frac{3}{\pi}} (ir + \omega)}{2 r^{3/2}}+ R(\omega) \frac{(1+i)(1+i\sqrt{3}) e^{\left(-\frac{1}{2} + \frac{i}{2}\right) \pi \omega - \frac{i \omega}{r}} \sqrt{\frac{3}{\pi}}  (-ir + \omega)}{2 r^{3/2}} \,.
\ee

Then, using the holographic dictionary yields for the retarded correlator (restoring the explicit dependence on $\beta$)
\be
G_R^{\rm defect}(\omega)=\frac{\omega^3 \left( (-i + \sqrt{3}) e^{\beta \omega / 2} + 2 i e^{i \beta \omega / 2} R(\omega) \right)}{(-2 - 2 i \sqrt{3}) e^{\beta \omega / 2} - 4 e^{i \beta \omega / 2} R(\omega)} = \frac{i \omega^3}{2} - e^{i\pi/6} R(\omega) \omega^3\, e^{-\fft{\beta}{2}(1-i)\omega}+\cdots\,.\label{eq: GR defect}
\ee
From this result, we can also read off the asymptotic quasinormal modes as
\be
\omega_q=\omega_{0q}+\frac{(1 - i) 2^{7/6} (-i + \sqrt{3}) C_{\fft{1}{6}}}{\pi \omega_{0q}^{4/3} \Gamma\left(-\frac{1}{6}\right)}\,,\quad \omega_{0q}=\fft{\pi}{\beta}\left(\fft{1 - i}{2} + 2(1 -  i) n + \frac{(1 + i) \log 3}{2\pi}\right)\,.
\ee

Eq.~(\ref{eq: GR defect}) is one of our main results for the retarded correlator of the displacement operator on the defect. We see that $G_R^{\rm defect}(\omega)$ displays the same exponentially suppressed terms $\sim e^{-\frac{\beta}{2}(1-i)\omega}$ at large frequency as in the bulk scalar case, though with different details of the reflection coefficient. By Fourier transforming \eqref{eq: GR defect}, the nonperturbative terms at large frequency give rise to 
\be
G_R^{\rm defect}(t)=\fft{3\sqrt{3}}{\pi (t-t_c)^4} - \fft{40\pi \Gamma\left(\fft{2}{3}\right)^2}{7\times 3^{\fft{1}{6}}\pi \Gamma\left(\fft{1}{6}\right)^2 (t-t_c)^{\fft{8}{3}}}+\ldots \,,
\label{defect-bouncing}
\ee
which shows the same bouncing singularity at $t_c=\beta/2(1+i)$ as in the correlator of local operators. Following the discussion in \ref{subsec: OPE-from-sing}, we find  that the corresponding large $n$ behavior of the retarded OPE coefficients is given by
\be
\hat{a}_{4n}^{R t}=4^{3+n}n^3 e^{-i n \pi}\fft{6}{\sqrt{3}\pi}\left(r_{3,\fft{3}{2}}+\frac{5\left(\tfrac{2}{3}\right)^{2/3}\pi^{7/3}\Gamma\!\left(\tfrac{2}{3}\right)^2\, r_{0,\,3/2}}
{7\,n^{4/3}\,\Gamma\!\left(\tfrac{1}{6}\right)^2\,\Gamma\!\left(\tfrac{8}{3}\right)}+\mathcal{O}(n^{-\fft{8}{3}})\right)\,,\label{eq: string a next}
\ee
where $r_{\alpha,\nu}$ is given in (\ref{r-alpha-nu}). 

We now see that although we are probing AdS$_2$/dCFT$_1$ on the defect, the operators that control the retarded correlator still have scaling dimensions $4n$ for integer $n$ (this will be transparent in the near-boundary analysis performed in the next subsection). The geometric reason is that the metric, although asymptotic to AdS$_2$, still exhibits a fall off in powers of $1/r^4$, inherited from the bulk black hole metric. Physically, this means that transverse gravitons interact with the fluctuations of the string worldsheet, forming bound states of open and closed strings. On the boundary, this implies that the defect displacement OPEs $D\times D$ contain operators that can be viewed as coming from the bulk-to-defect limit of multi stress-tensor operators. Schematically, we may denote such insertions along the Wilson line as $W T^n$ (the color trace structure of $W$ and $T^n$ is separate, but the $T^n$ operator is inserted along the Wilson line). Similar multi-trace operators arising in the defect OPEs are mentioned in \cite{Giombi:2018hsx}, however their role has not been discussed much in the literature, which mainly focused on the planar limit at zero-temperature. Our calculations provide an example where such operators play a significant role. In the next subsection, we extract this OPE systematically from the near-boundary expansion. 
%to our knowledge, there are no explicit computations in the literature, and we provide an example to demonstrate that they play a significant role. We will compute this OPE in later sections.

%\subsection{Defect OPEs and bouncing singularity}
\subsection{Near-boundary expansion and defect OPE}
\label{subsec: defect OPE}

We now study the defect OPEs along the Wilson line using the asymptotic OPE method \cite{Fitzpatrick:2019zqz,Fitzpatrick:2019efk,Li:2019tpf}.

We study the Euclidean equation
\be
f(r) \left( m(r)^2 r^2 \phi - r^4 f'(r) \partial_r \phi \right) - r^3 f(r)^2 \left( 2 \partial_r \phi + r \partial_r^2 \phi \right) - \partial_\tau^2 \phi=0\,,
\ee
Since we are studying AdS$_2$/dCFT$_1$, there is no transverse momentum or position to be integrated out or set to zero. Thus the near-boundary expansion of the solution is
\be
\phi(r,w)=\fft{r^2}{w^4} \Psi(r,w)\,,\quad \Psi(r,w)=\sum_{n=0}\fft{\mu^n g_{n}(w)}{r^{4n}}\,,\quad g_n(w)=\sum_{k=-2n}^{4n}Q_{nk}w^k \,.\label{eq: expansion phi}
\ee
where the prefactor is the bulk-to-boundary propagator in the pure AdS
\be
G_{b\partial}^{\rm AdS}\sim \fft{r^2}{(1+r^2\tau^2)^{2}}\,.
\ee
As we clearly see now, the solution obtained in this way organizes in powers of $\tau^{4n}$ in the retarded correlator, suggesting that the OPE coefficients correspond to multi-stress-tensors inserted along the Wilson line, schematically $W T^n$, as expained in the previous section. However, we note that the scaling dimension of the displacement operator is $\Delta=2$, which is an integer. As noted in \cite{Fitzpatrick:2019zqz}, the asymptotic OPE method has issues at integer $\Delta$, because multi-stress tensor operators (here multi-stress-tensors along the Wilson line) mix with ``two-particle" operators such as $D \partial_t^{2n} D$ (these are the analogs of the double-trace operators in the case of the two-point function of local single-trace operators dual to bulk scalars). As a result, the OPE coefficients develop integer poles $1/(\Delta-\text{integer})$. 

As we also explain in Appendix \ref{app: OPE x=0}, this is not an issue when computing the retarded correlator, because the analytic continuation gives a factor of $\sin(\pi\Delta-2n\pi)$ for each OPE coefficient (see \eqref{GR-defect-OPE}), canceling the divergence. If one starts directly with $\Delta=2$, there is still a solution of the form \eqref{eq: expansion phi}, but augmented by logarithmic terms $\log w/r$, which reduce to a constant after taking the commutator to compute the retarded correlator. The results are essentially the same as working with a general $\Delta$ first, and then taking the residues of the Euclidean OPE at the corresponding pole $\Delta=2$, which is known as the residue relation \cite{Li:2019tpf,Li:2020dqm}.

Therefore, practically, we keep $\Delta$ generic and, in the end, multiply $\sim\sin(\pi\Delta-2n\pi)$ and take $\Delta\rightarrow 2$. This is the conventional procedure for viewing retarded correlators expanded by the OPE in time as distributions. In other words, we can use the formula
\be
\hat{a}^{Rt}_{4n}=2\lim_{\Delta\rightarrow 2}Q_{n,4n}\pi^{4n}\sin\left(\Delta \pi-2n \pi\right)\,,\quad Q_{00}=\fft{3}{\pi}\,.
\ee
 For example, we find
\be
\hat{a}_{4}^{Rt}=-\fft{\pi^4}{2}\,,\quad \hat{a}_{8}^{R t}=\fft{11\pi^8}{24}\,,\quad a_{12}^{Rt}=-\fft{13231\pi^{12}}{161280}\,,\quad \hat{a}_{16}^{Rt}=\fft{3459557 \pi^{16}}{383201280}\,,\cdots\,.\label{eq: defect OPE}
\ee 

An alternative method is to directly start with the frequency space and solve $\phi_\omega$ by the expansion \eqref{eq: phi expand frequency} with $\nu=3/2$, as explained in subsection \ref{subsec: OPE omega}. By this procedure we find the result
\be
G_R^{\rm defect}(\omega)=\frac{i \omega^3}{2} - \frac{i}{2 \omega} + \frac{11 i}{\omega^5} - \frac{13231 i}{4 \omega^9} + \frac{17297785 i}{4 \omega^{13}} + \mathcal{O}(\omega^{-17})\,,\label{eq: large GR}
\ee
which can be verified to agree with \eqref{eq: defect OPE}.

In this case, we can also easily go to $n\sim 650$ and we find agreement with the asymptotic formula \eqref{eq: string a next} predicted by the WKB analysis, including the fitting of $1/n$ tails. In addition, we can also predict higher order tails such as $n^{-8/3}$ in a stable way
\be
& {\rm fit}[n]=\fft{\hat{a}_{4n}^{Rt}}{4^{3+n}n^3 e^{-i n \pi}\fft{6}{\sqrt{3}\pi}}-1 \sim  - \frac{1.5}{n} + \frac{0.309926}{n^{4/3}} + \frac{0.6875}{n^2} - \frac{0.344362}{n^{7/3}} + \frac{0.000404248}{n^{8/3}}\nn\\
& - \frac{0.09375}{n^3} + \frac{0.0741334}{n^{10/3}} - \frac{0.000112291}{n^{11/3}} + \frac{0.0804292}{n^4} + \frac{0.00513708}{n^{13/3}}+\cdots\,.
\ee
It would be interesting to match the subleading terms such as $n^{-8/3}$ with the WKB analysis, by extending the calculation in section \ref{subsec: defect bouncing WKB} to higher orders. We leave this to future work. 

\subsection{Numerical checks}
\label{subsec: numerical}

In this subsection, we perform numerical checks of our analytic results, in particular the prediction for the nonperturbative terms we obtained in eq.~(\ref{eq: GR defect}). 

\begin{figure}[h]
    \centering
    % --- First Subfigure ---
    \begin{subfigure}[b]{0.5\textwidth}
        \centering
        % Replace 'example-image-a' with your filename
        \includegraphics[width=\textwidth]{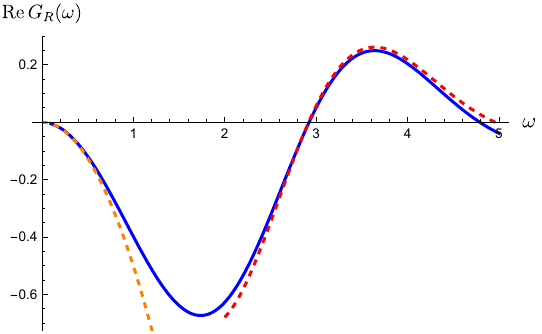}
        \caption{}
        \label{fig: sub1}
    \end{subfigure}
    \hfill % Adds flexible space between images to push them apart
    % --- Second Subfigure ---
    \begin{subfigure}[b]{0.48\textwidth}
        \centering
        % Replace 'example-image-b' with your filename
        \includegraphics[width=\textwidth]{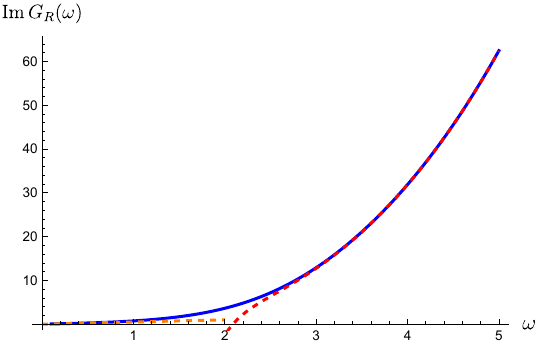}
        \caption{}
        \label{fig: sub2}
    \end{subfigure}
    \caption{Numerical comparison for the retarded correlator of the displacement operator on a thermal Wilson line. The solid blue line is obtained from the numerical solution. The red dashed line shows the analytic large-$\omega$ result, obtained by combining \eqref{eq: GR defect} and \eqref{eq: large GR}, and the orange dashed line is the analytic small-$\omega$ result given in \eqref{eq: GR small}.}
    \label{fig: numerical checks}
\end{figure}

\begin{figure}[t]
    \centering
    \includegraphics[width=0.7\textwidth]{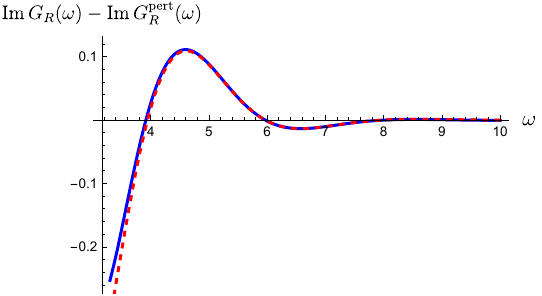}
    \caption{Comparison between the analytic prediction for the nonperturbative part of the spectral density (red dashed line) and the numerical solution with the perturbative part subtracted (blue solid line).}
    \label{fig: numerical check 2}
\end{figure}

More specifically, we numerically solve \eqref{eq: eq string phi} with infalling condition for the massive modes by transforming it into a Heun equation, as shown in Appendix \ref{app: Heun}, and evaluate the retarded correlator. We then compare the numerical solutions with our analytic results for both small frequency \eqref{eq: GR small} and large frequency, which is given by the perturbative part \eqref{eq: large GR} plus the nonperturbative corrections we obtained from the WKB method (the second term in \eqref{eq: GR defect}, with the reflection coefficient given in \eqref{eq: defect R reflection}). The comparison is shown in Fig.~\ref{fig: numerical checks}. Note that at large frequencies the real part of the retarded correlator is purely nonperturbative, and the oscillatory behavior of the numerical solution matches well with the analytic prediction in  \eqref{eq: GR defect}.   
%We find that the nonperturbative corrections at large $\omega$ are significant for producing the oscillatory behavior in the real part of the retarded correlator. 
The imaginary part of the retarded correlator, i.e. the spectral density, is dominated at large $\omega$ by the vacuum contribution. In order to isolate the non-perturbative oscillatory behavior in a more transparent way, in Fig \ref{fig: numerical check 2} we compare the prediction of \eqref{eq: GR defect} to the numerical solution after subtracting from it the perturbative part given by \eqref{eq: large GR}.  
%can also be compared with numerical solutions in a more transparent way by subtracting the nonperturbative corrections, as shown in Fig \ref{fig: numerical check 2}.  

\subsection{On the massless modes}
\label{subsec: massless}

Let us now consider the massless modes that describe S$^5$ and correspond to SYM scalars. In this case, the equation \eqref{eq: eq string phi} with $m=0$ can be solved exactly. To see this, note that by changing the radial variable to the tortoise coordinate defined by
\be 
\frac{dr_s}{dr} = \frac{1}{r^2f(r)}
\label{tortoise-rs}
\ee 
the induced metric of the worldsheet takes the conformally flat form
\be 
ds^2 = -r^2 f(r)dt^2 + \frac{dr^2}{r^2f(r)} = r^2 f(r) \left(-dt^2+dr_s^2\right)
\ee 
Since we have a 2d massless field, the wave equation in tortoise coordinates then reduces to the simple flat space wave equation
\be 
\left(-\partial_t^2+\partial_{r_s}^2\right)\phi^{m=0}(t,r_s)=0
\ee 
The solution with infalling boundary condition is $\phi^{m=0}(t,r_s)=e^{-i\omega(t+r_s)}$, or $\phi^{m=0}_{\omega}(r)=e^{-i\omega r_s(r)}$. Expressing this result in terms of the original radial coordinate by explicitly solving (\ref{tortoise-rs}), 
the exact solution is then (restoring the dependence on $r_h$)
\be
\phi_\omega^{m=0}(r)=  \left(\fft{r+ir_h}{r-ir_h}\right)^{\fft{\omega}{4r_h}}\left(\fft{r-r_h}{r+r_h}\right)^{-i \fft{\omega}{4r_h}}\,.
\ee
Using the dictionary \eqref{eq: holographic dic} yields straightforwardly
\be
G_R^{m=0}(\omega)=\fft{i \omega}{2}\,.\label{eq: GR omega m0}
\ee
Note that in this case the retarded correlator does not display a bouncing singularity. 

Let's also comment on how to understand this result from OPE perspective. We can easily see that there is no other contributions in the retarded OPE but the identity operator (all $Q$ except for $Q_{00}$ are identically zero). We thus perform the previous regulating procedure and find
\be
G_R^{m=0}(t)= \lim_{\Delta\rightarrow 1} \fft{-\sin(\Delta \pi)}{\pi}\fft{\theta(t)}{ t^{2\Delta}}\sim - \fft{\delta'(t)}{2}\,,
\ee
where we understand such a limit as a distribution using 
\be
x^a \theta(x)=\fft{(-1)^{n-1}\partial_x^{n-1}\delta(x)}{(n-1)!}\fft{1}{a+n}+\cdots\,.
\ee
This indeed agrees with the Fourier transform of \eqref{eq: GR omega m0}.

In this case, the exact Wightman function and the Euclidean thermal correlator can also be obtained via the fluctuation-dissipation theorem together with a Fourier transform.
\be
G^{m=0,>}(t)=-\fft{\pi^2}{\beta^2} \fft{1}{\left(\sinh\left(\fft{\pi t}{\beta}\right)\right)^2}\,,\quad G_{E}^{m=0}(\tau)=\fft{\pi^2}{\beta^2} \fft{1}{\left(\sin\left(\fft{\pi \tau}{\beta}\right)\right)^2}\,.
\ee
To understand the Euclidean OPE, we study the small-$\tau$ expansion, from which we find
\be
a_2=\fft{\pi^2}{3}\,,\quad a_{4}= \fft{\pi^4}{15}\,,\quad a_{6}=\fft{2\pi^6}{189}\,,\quad a_{8}=\fft{\pi^8}{675}\,, \cdots\,.
\ee
%The absence of logarithmic terms in the Euclidean correlator indicates that multi-stress-tensor contributions are absent. 
%As a result, 
It would be interesting to clarify how these Euclidean OPE coefficients are generated by the mixing between double-trace-like operators of the form $\Phi \partial^{2n}\Phi$ with $\Delta' = 2 + 2n$ and multi-stress-tensor operators, all of which are killed by the retarded commutator. 
%(therefore, they are not correctly captured by the asymptotic OPE method $Q$, which only computes the multi-stress-tensor sector).

\section{Universality, EFT and factorization}
\label{sec: universality}

In this section we comment on the universality suggested by our computations. The discussion below applies in the same way for the case of bulk scalar and defect correlators. 
We argue that the bouncing singularity and the associated multi-stress-tensor OPE data 
encode a universal high-frequency structure of the retarded correlators, independent of 
the detailed interior geometry. We further propose a factorization structure that 
separates this universal contribution from state-dependent physics, and outline a 
holographic setup where deviations from universality could be detected.

\subsection{Heuristic arguments and EFT interpretation}

A remarkable feature emerging from our analysis is the precise agreement between two 
seemingly unrelated approaches:

\begin{itemize}
\item The WKB proposal \cite{Afkhami-Jeddi:2025wra}, which probes the near-singularity region of the 
black hole interior and directly predicts nonperturbative tails of the retarded Green's function and, via a Fourier transform, the bouncing singularity $G_R(t,p=0) \sim (t-t_c)^{-2\nu-1}$. 

\item The asymptotic OPE method \cite{Fitzpatrick:2019zqz}, which probes only the near-boundary region 
and determines the large-$n$ behavior of multi-stress-tensor OPE coefficients.
\end{itemize}
Both methods give the same nonperturbative tail and therefore the same bouncing 
singularity. This agreement is highly nontrivial. The WKB analysis explicitly 
depends on the existence of a horizon and singular interior geometry, whereas 
the asymptotic OPE method is blind to the deep bulk and does not know whether the 
interior contains a black hole, a neutron star, or any other compact object.

A sharper way to state the puzzle is the following. The WKB computation directly 
evaluates thermal retarded correlators. By contrast, the asymptotic OPE method 
more generally computes heavy–light–light–heavy four–point functions without 
assuming thermality. Nevertheless, the same OPE data appear. This is reminiscent 
of the thermalization of heavy states \cite{Srednicki:1999bhx,DAlessio:2015qtq,Lashkari:2016vgj,Collier:2019weq,Delacretaz:2020nit,Karlsson:2022osn,Karlsson:2021duj,Dodelson:2022yvn,Esper:2023jeq}
\be
\langle H|[\mathcal{O}(x),\mathcal{O}(0)]\theta(t)| H\rangle
\sim 
\langle[\mathcal{O}(x),\mathcal{O}(0)]\theta(t)\rangle_\beta .
\ee
Our results then suggest the following statement: {\it The bouncing singularity is a distinctive feature of black holes, yet it is encoded in universal high-frequency OPE data of the retarded correlator.}
%show that this relation extends into the high-frequency regime, which thus indicates the following statement:

%\begin{itemize}
%\item[] {\bf The bouncing singularity is a distinctive feature of black holes, yet it is encoded in universal high-frequency OPE data of the retarded correlator.}
%\end{itemize}

To elaborate on this point further, let us emphasize that the asymptotic OPE method is essentially a framework of long-distance EFT in AdS. Around the AdS energy scale we have \cite{Polchinski:2001tt,Heemskerk:2009pn}
\be
\fft{r}{R_{\rm AdS}^2}\sim \mu_{\rm FT}\,,
\ee
where $r$ denotes the length scale probed by bulk excitations, and $\mu_{\rm FT}$ is the corresponding energy scale on the field theory side. Therefore, the long-distance EFT in AdS captures UV CFT data, namely the OPE coefficients, from the boundary perspective. The universality thus originates from universal UV features of the OPE in the CFT, regardless of the state probed by the operators. %In asymptotic AdS, we thus expect a factorization formula at high-frequency as
It is thus natural to expect a factorization formula of the form
\be
G_R(\omega)\sim G_R^{\rm FZ}(\omega) G_R^{\rm NZ}(\omega)\,,
\ee
where $G_R^{\rm FZ}$ captures the long-distance physics perturbatively in $1/\omega$ (FZ for ``far-zone'' schematically) and $G_R^{\rm NZ}$  knows more details of the states (NZ for ``near-zone''). Similar factorization formulas have been established for asymptotically flat spacetime in the low-frequency limit \cite{Ivanov:2022qqt,Bautista:2023sdf,Ivanov:2024sds,Caron-Huot:2025tlq,Saketh:2024juq,Ivanov:2026icp}, see also \cite{Caron-Huot:2025she,Caron-Huot:2025hmk} for the discussions in generic asymptotic AdS.

We can indeed see this structure in the planar black hole case using the formally exact retarded correlator \eqref{eq: G_R exact}. Although for planar black hole case, the instanton parameter $t_0=1/2$ is not small enough to make this formula \eqref{eq: G_R exact} practically useful, it indeed makes the factorization structure manifest at large-$\omega$ by noting that the Matone relation gives $a\sim \mathcal{O}(1)\times  i \omega$ (numerics suggest that $\mathcal{O}(1)\sim 1/4$). $\sigma=-1$ sector is then exponentially subdominant to $\sigma=1$ sector in both numerator and denominator by $e^{-\beta/2 \omega}$. This suggests a factorization formula 
\be
 G_R^{\rm BH-FZ}&=\left[r_h^{4a_1}e^{-\partial_{a_1} F} \fft{\mathcal{M}_{-1,1}(a_{t_0},a;a_0)\mathcal{M}_{-1,1}(a,a_1;a_{\infty})}{\mathcal{M}_{-1,1}(a_{t_0},a;a_0)\mathcal{M}_{-1,-1}(a,a_1;a_{\infty})}\right]_{\rm pert}\sim -\fft{e^{-i\pi\nu}}{\sin \pi\nu}\omega^{2\nu+1}\left(1+\cdots\right)\,,\nn\\
  G_R^{\rm BH-NZ}&=\left(1+ \left[r_h^{4a_1}e^{-\partial_{a_1} F} \fft{\mathcal{M}_{-1,1}(a_{t_0},a;a_0)\mathcal{M}_{-1,1}(a,a_1;a_{\infty})}{\mathcal{M}_{-1,1}(a_{t_0},a;a_0)\mathcal{M}_{-1,-1}(a,a_1;a_{\infty})}\right]_{\rm nonpert}\right)\nn\\
  & \times \fft{1+\fft{\mathcal{M}_{-1,-1}(a_{t_0},a;a_0)\mathcal{M}_{1,1}(a,a;a_{\infty})}{\mathcal{M}_{-1,1}(a_{t_0},a;a_0)\mathcal{M}_{-1,1}(a,a;a_{\infty})}}{1+\fft{\mathcal{M}_{-1,-1}(a_{t_0},a;a_0)\mathcal{M}_{1,-1}(a,a_1;a_{\infty})}{\mathcal{M}_{-1,1}(a_{t_0},a;a_0)\mathcal{M}_{-1,-1}(a,a_1;a_{\infty})}}
 \sim 1+ e^{-\fft{\beta}{2}\omega} K(\omega) +\cdots\,,\label{eq: BH GR all}
\ee
We indeed see the expected factorization structure, where $G_R^{\rm BH-FZ}$ encodes the asymptotic results expanded in $1/\omega$, which is universal. Our computation further suggests that the universality of the multi–stress–tensor OPEs persist asymptotically as $n\rightarrow\infty$, however, whether they can be resummed to provide the nonperturbative tail and complex time singularity depends on the the interior physics. It turns out that high-frequency structure is more appropriately probed from the formally exact retarded correlator from Heun connections using the $s$-channel Virasoro block \cite{Jia:2024zes} (rather than the $t$-channel in \eqref{eq: G_R exact}), where this factorization structure can be put on a more rigorous footing and can indeed show that $G_R^{\rm BH\text{-}FZ}$ is universal and encodes only OPE data, while $G_R^{\rm BH\text{-}NZ}$ captures nonperturbative corrections \cite{Jia:2026ryl}, making this structure practically useful. %thereby providing the nonperturbative tail and the bouncing singularity. 

Therefore, we conjecture a factorization formula for any compact objects
\be
G_R(\omega)\sim G_R^{\rm BH-FZ}(\omega) \hat{G}_R^{\rm NZ}(\omega)\,,
\ee
where $G_R^{\rm BH-FZ}$ denotes the retarded correlator computed for black holes via the asymptotic OPE method. On the other hand, $\hat{G}_R^{\rm NZ}$ denotes the remainder that depends on additional details of the interior physics.

From the boundary perspective, we consider the OPE expansion in general, schematically (for a 1D defect CFT, or for $p=0$ in the case of a bulk primary operator).
\be
\langle H|[\mathcal{O}(t),\mathcal{O}(0)]\theta(t)|H\rangle\sim \fft{1}{t^{2\Delta}}\sum_{\Delta'} a_{\Delta'}^{R t}\, t^{\Delta'} \langle H|\mathcal{O}'(0)|H\rangle\,.
\ee
$G_R^{\rm BH\text{-}FZ}$ universally determines the multi–stress–tensor OPE coefficients $a_{dn}^{Rt}$ for all $n$, while $\hat{G}_R^{\rm NZ}$ encodes details of $\langle H|\mathcal{O}'|H\rangle$ and further contributions completely beyond the OPE limit. Specifically, only for thermal states, where $\langle \mathcal{O}'\rangle_\beta \sim (1/\beta)^{\Delta'}$, we know how to appropriately resum the OPE to obtain the bouncing singularity. However, in general for other heavy states, the validity of this resummation is not guaranteed because in principle we might have $\langle H| \mathcal{O}'| H \rangle\sim (1/\beta_{\rm eff})^{\Delta'}\left(1+\mathcal{O}(e^{-\beta_{\rm eff}/L(\Delta')})\right)$, where $L$ represents other scale in the system that may depend on $\Delta'$. This implies that the retarded correlator is not generally analytic in real time beyond some finite time, corresponding to when the wave propagates and hits the surface of a compact object, because its short-time OPE agrees with that of a black hole while the full correlator is very different.\footnote{We thank Simon Caron-Huot for pointing out this implication.} Thermality appears to eliminate this real time nonanalyticity by maximizing absorption, with the bouncing singularity emerging as a consequence.

It is worth noting that the universality and factorization structures discussed here are all based on holographic pictures involving compact objects, namely heavy operators in the CFT. It is unclear whether our discussion remains valid beyond holographic CFTs.

\subsection{A possible route to test: holographic neutron stars}

Let us now provide a more concrete setup to test the heuristic arguments in the previous subsection. However, we leave the details of this exercise to future work.

It is instructive to consider holographic neutron stars \cite{deBoer:2009wk} (in the planar limit)
\be
ds^2=e^{2\sigma(r)}r^2\left(1-\fft{M(r)}{r^{d}}\right)+ \fft{dr^2}{r^2\left(1-\fft{M(r)}{r^d}\right)}+r^2 dx^2\,,
\ee
which satisfies the equation (in $d=4$)
\be
M'(r)=\fft{2}{3}r^3 \rho(r)\,,\quad \sigma'(r)=\fft{r^3(p(r)+\rho(r))}{3(r^4-M(r))}\,.
\ee
$(\rho(r),p(r))$ denote the density and pressure of the Fermi gas
\be
& T_{\mu\nu}=(\rho+p)u_\mu u_\nu+p g_{\mu\nu}\,,\quad \frac{(p(r) + \rho(r)) f'(r)}{2 f(r)} + p'(r) + \frac{(p(r) + \rho(r))(1 + r \sigma'(r))}{r}=0\,,\nn\\
& f(r)=1-\fft{M(r)}{r^{4}}\,.
\ee
It is natural to consider the boundary condition $M(0)=0$, and the total mass is $M(R)$, where $R$ is the radius of the neutron star
\be
1-\fft{M(R)}{R^4}=\left(\fft{\epsilon_F}{m_f}\right)^2\,,
\ee
with Fermi energy $\epsilon_F$ and fermion mass $m_f$. We can consider either probing a neutron star using scalar fields in asymptotic AdS$_5$, or studying neutron stars on the worldsheet, which corresponds to computing heavy-light four-point functions of bulk primary scalars or on the line defect.

The crucial feature of the neutron star is that the metric outside the star is identical to the Schwarzschild black hole
\be
\chi(r\geq R)=0\,,\quad M(r\geq R)=M(R)\,.
\ee
Therefore, without performing any explicit computations, we can summarize that applying the asymptotic OPE method to a neutron star yields exactly the same multi–stress–tensor OPE coefficients $a_{4n}^{R t}$ as for a black hole, for all $T^n$ or $WT^n$. We can thus obtain $G_R^{\rm BH-FZ}$. However, this is not the end of the story. The interior structure of neutron stars differs from that of black holes, and the boundary conditions at the surface of a neutron star are also different from the infalling boundary condition at a black hole horizon. The full $G_R$ in a heavy state corresponding to a neutron star can only be computed by incorporating detailed information from the surface and interior, which is required to determine $\hat{G}_R^{\rm NZ}$, or at least, in the OPE limit, $\langle H|T^n|H\rangle$. This is the key contribution that drives deviations from the universality of heavy state thermalization, and is expected to be exponentially small by the eigenstate thermalization hypothesis \cite{Lashkari:2016vgj}.

\section{Summary}
\label{sec: summary}

In this paper we studied thermal retarded two-point functions in holographic CFTs, focusing on the high-frequency regime. We reviewed in detail the case of bulk scalars dual to single-trace local operators, and further studied retarded correlators of displacement operators on Wilson lines at finite temperature.

For bulk scalar operators at zero spatial momentum, we revisited the large-frequency WKB analysis with infalling boundary conditions at the horizon \cite{Afkhami-Jeddi:2025wra}. By performing a careful near-origin analysis and matching onto the WKB phase, we reproduced the exponentially suppressed correction of order $e^{-\beta \omega/2}$. In the time domain, this gives rise to a complex singularity located at $t_c = \frac{\beta}{2}(1+i)$, known as the bouncing singularity. We showed explicitly how this structure emerges from the high-frequency behavior of the bulk solution. Independently, we applied the asymptotic OPE method \cite{Fitzpatrick:2019zqz}, both in time domain and directly in frequency space. This method relies only on the near-boundary expansion and does not use any information about the black hole interior. We computed the large-$n$ asymptotics of the multi-stress-tensor OPE coefficients $T^n$, including subleading fractional corrections of order $1/n^{4/3}$, and demonstrated that their resummation reproduces precisely the same nonperturbative tail and complex-time singularity obtained from the WKB analysis.

We then extended the analysis to thermal Wilson lines, described holographically by a static string in the planar AdS black hole background. We computed the retarded correlator of displacement operators using both the WKB method \cite{Afkhami-Jeddi:2025wra} and the asymptotic OPE method \cite{Fitzpatrick:2019zqz} adapted to the present case of the asymptotically AdS$_2$ string dual to the line defect. In this case, the bouncing singularity is controlled by towers of defect operators of the schematic form $W T^n$ (multi-stress tensor operators ``attached" to the Wilson line). Similarly as in the bulk scalar case, we find that the two approaches are precisely consistent with each other, including subleading large-$n$ corrections.

The agreement between a method that probes propagation toward the interior and a method that uses only near-boundary data is surprising and indicates universal features of thermalization in the high-frequency regime. %The multi-stress-tensor sector determines a universal asymptotic structure of the retarded correlator. 
We proposed a factorization formula at high-frequency, separating a universal component controlled by multi-stress-tensors from a non-universal, state-dependent component. While the universal sector fixes the high-frequency asymptotics, whether this data can be resummed into a bouncing singularity depends on the non-universal sector. In this sense, the existence of the bouncing singularity reflects both universal multi-stress-tensor OPE and one-point functions of multi-stress-tensors in a given state. This perspective also suggests a concrete setting to test deviations from black hole universality, for instance in holographic neutron-star backgrounds, where the universal asymptotics persists but the resummation structure may differ.

Several future directions naturally follow from our results:

\begin{enumerate}

\item {\bf Compact objects and factorization.}  
It would be important to test the proposed factorization structure in backgrounds other than the planar black hole, such as holographic neutron stars, boson stars, or other compact objects. This can be studied both for bulk scalar correlators and for Wilson line correlators. Such setups provide a natural playground to test whether the universal high-frequency asymptotic OPE persists while the non-universal sector indeed changes, and to understand more precisely when the asymptotic data can or cannot be resummed into a bouncing singularity. This might shed light on understanding thermalization at high-frequency.

\item {\bf Higher-derivative and string corrections.}  
Another natural direction is to include $\alpha'$ or higher-curvature corrections, for example by adding Gauss-Bonnet terms in the bulk theory, or by introducing non-minimal couplings of particles in the bulk \cite{Fitzpatrick:2020yjb}. Since the bouncing singularity is sensitive to the high-frequency behavior of the wave equation, it is expected to probe such corrections. Studying these effects using both the WKB method and the asymptotic OPE method may clarify how the universal structure is modified.

\item {\bf More general defect theories.}  
In the Wilson line setup, the holographic description involves a fundamental string. One may consider a more general string action in AdS, including higher-derivative corrections to the Nambu-Goto action, as in flux-tube effective string theory \cite{Polchinski:1991ax,Dubovsky:2012sh,Aharony:2013ipa,Luscher:2004ib,Albert:2026fqj} but now in AdS \cite{Gabai:2025hwf,Gabai:2026myo}. These corrections should modify the defect OPE data and provide another controlled way to study the universality structure with a line defect in CFT.

\item {\bf Other geometries.}  
It would also be interesting to generalize the analysis to other backgrounds, such as spherical black holes for both bulk scalar operators and defect operators along a Wilson line. In these cases the analytic structure of the correlators and the quasinormal spectrum differ from the planar case, see for example \cite{Dodelson:2023nnr,Jia:2026pmv}, and it would be useful to understand how the factorization picture and the bouncing singularity are modified.

\item {\bf Exact WKB and the factorization formula.}  
On a more formal level, it would be interesting to understand the proposed factorization structure directly from the wave equation, for example using exact WKB methods along the lines of \cite{Jia:2025jbi,Jia:2026ryl}. Such an approach may clarify how the universal and non-universal sectors arise from the monodromy data and how the bouncing singularity is encoded in the exact analytic structure.

\item {\bf Higher-point functions.}  
It is natural to consider higher-point correlators, such as four-point functions. The Wilson line setup may be particularly useful, since the string description provides explicit interaction vertices by expanding the string action in powers of fluctuations, and there is no spatial momentum to keep track of. It would be interesting to understand whether higher-point functions can probe the black hole interior more directly and how the factorization structure extends beyond two-point functions.

\item {\bf Small-frequency regime.}  
Another open question is whether the black hole singularity can also be probed at small frequency. This is related to recent work where the singularity appears in one-point functions, corresponding to the zero-frequency limit \cite{Grinberg:2020fdj}. Effective thermal field theory \cite{Benjamin:2023qsc} may help clarify the structure in this case.

\item {\bf Other observables and non-equilibrium states.}  
Instead of two-point functions of local operators, one may consider other observables such as energy-energy correlators (EEC) \cite{Hofman:2008ar,Moult:2025nhu}, where factorization properties are well understood in vacuum CFT. It would be interesting to see whether in thermal or time-dependent states such observables show deviations that carry information about the interior geometry. More generally, studying out-of-equilibrium states may help identify which correlators (retarded or Wightman) are most sensitive to nontrivial interior structure.

\end{enumerate}

\acknowledgments We thank Ahmed Almheiri, Simon Caron-Huot, Matthew Dodelson, Ping Gao, Hewei Jia, Hong Liu, Andrei Parnachev, Mukund Rangamani and David Simmons-Duffin for helpful discussions. This work supported by Simons Foundation grant No. 917464 (Simons Collaboration on Confinement and QCD Strings).

\newpage

\appendix

\section{Wave equations as Heun equations}
\label{app: Heun}

In this Appendix, we show that the wave equations \eqref{eq: bulk scalar eq p} and \eqref{eq: eq string phi} can be transformed into Heun form. This allows the retarded correlator to be written formally in terms of NS functions \cite{Aminov:2020yma,Bonelli:2022ten,Bonelli:2021uvf,Bianchi:2021mft,Bianchi:2021xpr,Fioravanti:2021dce,Consoli:2022eey,Dodelson:2022yvn,Aminov:2023jve,Bautista:2023sdf,Jia:2024zes}.

The bulk scalar wave equation in AdS$_5$ can be cast in Heun form, as shown in \cite{Dodelson:2022yvn}. We begin with \eqref{eq: bulk scalar eq p} and consider the transformation
\be
\phi_{\omega p}=\left(\fft{1-z}{z(2z-1)}\right)^{\fft{1}{2}}\chi(z)\,,\quad \fft{r_h^2}{r^2}=\fft{1-z}{z}\,.
\ee
We then find that the equation becomes a Heun equation
\be
\chi''(z)+\left(\frac{1/4 - a_0^2}{z^2} + \frac{1/4 - a_1^2}{(z-1)^2} - \frac{1/2 + u - a_{t_0}^2 - a_0^2 - a_1^2 + a_\infty^2}{z(z-1)} + \frac{1/4 - a_{t_0}^2}{(z - t_0)^2} + \frac{u}{z(z - t_0)}\right)\chi(z)=0\,.\label{eq: Heun}
\ee
where
\be
& t_0=\fft{1}{2}\,,\quad a_1=\fft{\sqrt{m^2+4}}{2}=\fft{\nu}{2}\,,\quad a_{t_0}=\fft{i \sqrt{\omega^2-p^2}}{4 r_h}\,,\quad a_0=0\,,\quad a_{\infty}=\fft{\omega}{4r_h}\,,\nn\\
&  u=\fft{\omega^2-p^2-2m^2 r_h^2-8r_h^2}{8r_h^2}\,.
\ee

For massive fluctuations on the string worldsheet embedded in AdS$_5$, we start with \eqref{eq: eq string phi} and now perform the transformation
\be
\phi_{\omega}=\left(\fft{z}{(1-z)(2z-1)^2}\right)^{\fft{1}{4}}\chi(z)\,,\quad \fft{r_h^2}{r^2}=\fft{1-z}{z}\,.
\ee
Although there is a $r$-dependent mass in the wave equation, we still find that it can be put in the standard Heun's equation form \eqref{eq: Heun}, where the parameters are identified as
\be
t_0=\fft{1}{2}\,,\quad a_1=\fft{3}{4}\,,\quad a_{t_0}=\fft{i\omega}{4r_h}\,,\quad a_0=\fft{1}{4}\,,\quad a_{\infty}=\fft{\omega}{4r_h}\,,\quad u=\fft{\omega^2-4r_h^2}{8r_h^2}\,.
\ee

Using the connection formulas of Heun equation \cite{Bonelli:2022ten}, the retarded thermal correlator can be formally written as (see \cite{Dodelson:2022yvn})
\be
G_R(\omega,p)=r_h^{4a_1} e^{-\partial_{a_1}F} \frac{\sum_{\sigma'=\pm}\mathcal{M}_{-\sigma'}(a_{t_0},a;a_0)\mathcal{M}_{(-\sigma')+}(a,a_{1};a_{\infty})t^{\sigma' a}e^{-\frac{\sigma'}{2}\partial_a F}}{\sum_{\sigma=\pm}\mathcal{M}_{-\sigma}(a_{t_0},a;a_0)\mathcal{M}_{(-\sigma)-}(a,a_{1};a_{\infty})t^{\sigma a}e^{-\frac{\sigma}{2}\partial_a F}}\,,\label{eq: G_R exact}
\ee
where $a$ satisfies the Matone relation \cite{Matone:1995rx,Flume:2004rp}
\be
u=-a^2+a_{t_0}^2-\fft{1}{4}+a_0^2+t_0 \partial_{t_0} F\,,
\ee
We also have
\be
\mathcal{M}_{\theta\theta'}(\alpha_0,\alpha_1;\alpha_2)=\fft{\Gamma\left(-2\theta'\alpha_1\right)\Gamma\left(1+2\theta'\alpha_0\right)}{\Gamma\left(\fft{1}{2}+\alpha_2+\theta \alpha_0-\theta' \alpha_1\right)\Gamma\left(\fft{1}{2}-\alpha_2+\theta \alpha_0-\theta' \alpha_1\right)}\,.
\ee
In this formula, $F$ is the instanton part of the NS free energy, we refer readers to \cite{Dodelson:2022yvn} for more details.

\section{Useful integrals}
\label{app: integrals}

In this Appendix, we discuss some useful integrals for evaluating the $x\rightarrow\infty$ limit of the near-origin solutions. We discuss some examples to show the idea, which can be straightforwardly generalized.

For bulk scalars, we have to perform the integral such as
\be
& \int_0^x dx' x'^a H_0^{(1)}(x')H_0^{(2)}(x')=\frac{x^{1+a} \, {}_2F_3\left(\frac{1}{2}, \frac{1+a}{2}; 1, 1, \frac{3+a}{2}; -x^2\right)}{1+a} 
+ \frac{x^{1+a} \Gamma\left(\frac{1+a}{2}\right) {}_2F_3\left(\frac{1}{2}, \frac{1+a}{2}; 1, 1, \frac{3+a}{2}; -x^2\right)}{2 \Gamma\left(\frac{3+a}{2}\right)} 
\nn\\
& + \frac{1}{\sqrt{\pi}} G_{3, 5}^{3, 1}\left(x^2 \middle| \begin{array}{c} 1, \frac{2+a}{2}, \frac{2+a}{2} \\ \frac{1+a}{2}, \frac{1+a}{2}, \frac{1+a}{2}, 0, \frac{2+a}{2} \end{array} \right)\,.
\ee
Our goal is to analyze its large-$x$ behavior. We choose a particular sector to kill all oscillating terms such that we end up with asymptotic behaviors expected by WKB. We consider the integral representations
\be
&\,_2F_3(a_1,a_2;b_1,b_2,b_3;x)=\int \fft{ds}{2\pi i}\frac{(-x)^s \Gamma(-s) \Gamma(s + a_1) \Gamma(s + a_2) \Gamma(b_1) \Gamma(b_2) \Gamma(b_3)}{\Gamma(a_1) \Gamma(a_2) \Gamma(s + b_1) \Gamma(s + b_2) \Gamma(s + b_3)}\,,\nn\\
&  G_{3, 5}^{2, 2}\left(x^2 \middle| \begin{array}{c} a_1, a_2, a_3 \\ a_4, a_5, a_6, a_7, a_8 \end{array} \right)=\int \fft{ds}{2\pi i}\frac{x^{-2s} \Gamma(1 - s - a_1) \Gamma(1 - s - a_2) \Gamma(s + a_4) \Gamma(s + a_5)}{\Gamma(s + a_3) \Gamma(1 - s - a_6) \Gamma(1 - s - a_7) \Gamma(1 - s - a_8)}\,.
\ee
We deform the contour to pick up negative poles for hypergeometric functions, and non-negative poles for MeijerG functions, which are required by having nice asymptotic structures. For example, for $ {}_2F_3\left(\frac{1}{2}, \frac{1+a}{2}; 1, 1, \frac{3+a}{2}; -x^2\right)$, we pick up poles at $s=-1/2(1+a), -1/2, -3/2,\cdots$ and find
\be
 {}_2F_3\left(\frac{1}{2}, \frac{1+a}{2}; 1, 1, \frac{3+a}{2}; -x^2\right)\sim \frac{1+a}{8(a-2)\pi x^3} - \frac{1+a}{a \pi x} - \frac{(1+a) x^{-1-a} \Gamma\left(-\frac{a}{2}\right) \Gamma\left(\frac{1+a}{2}\right)}{2 \sqrt{\pi} \Gamma\left(\frac{1-a}{2}\right)^2}+\cdots \,.
\ee
For $ G_{3, 5}^{3, 1}\left(x^2 \middle| \begin{array}{c} 1, \frac{2+a}{2}, \frac{2+a}{2} \\ \frac{1+a}{2}, \frac{1+a}{2}, \frac{1+a}{2}, 0, \frac{2+a}{2} \end{array} \right)$, it is sufficient to just take the pole at $s=0$
\be
 G_{3, 5}^{3, 1}\left(x^2 \middle| \begin{array}{c} 1, \frac{2+a}{2}, \frac{2+a}{2} \\ \frac{1+a}{2}, \frac{1+a}{2}, \frac{1+a}{2}, 0, \frac{2+a}{2} \end{array} \right)\sim -\frac{\Gamma\left(\frac{1+a}{2}\right)^3}{\Gamma\left(-\frac{a}{2}\right) \Gamma\left(\frac{2+a}{2}\right)^2}+\cdots \,.
\ee
Other integrals follow similar analysis, including those required for the analysis of the wave equation for the massive scalars on the string worldsheet.

\section{More fits for scalar thermal retarded correlator with $p=0$}
\label{app: more fits p=0}

In this Appendix, we provide fits for more $\Delta$ to verify \eqref{eq: a4n asymp} and further predict new coefficients in front of, e.g., $n^{-8/3}$, for various values of $\Delta$. 

We follow \cite{Ceplak:2024bja} to choose $\Delta=31/20,51/20,71/20,\cdots 131/20$. In addition to \eqref{eq: fits bulk scalar some}, we further have:
\be
\Delta=\fft{71}{20}\,,\quad {\rm fit}[n]& =-\frac{1.58875}{n} + \frac{0.394119}{n^{4/3}} + \frac{0.785438}{n^{2}} - \frac{0.472888}{n^{7/3}} + \frac{0.00113924}{n^{8/3}} - \frac{0.121474}{n^{3}} \nn\\
& + \frac{0.119269}{n^{10/3}} - \frac{0.000417561}{n^{11/3}} + \frac{0.10663}{n^{4}} + \frac{0.00589396}{n^{13/3}}\,,\nn\\
\Delta=\fft{91}{20}\,,\quad {\rm fit}[n]&=-\frac{3.88875}{n} + \frac{1.2766}{n^{4/3}} + \frac{5.74644}{n^{2}} - \frac{4.46793}{n^{7/3}} + \frac{1.53372}{n^{8/3}} - \frac{4.00428}{n^{3}} \nn\\
& + \frac{5.5777}{n^{10/3}} - \frac{4.08972}{n^{11/3}} + \frac{1.65682}{n^{4}} - \frac{2.87126}{n^{13/3}}\,,\nn\\
 \Delta=\fft{111}{20}\,,\quad {\rm fit}[n]&=-\frac{7.18875}{n} + \frac{2.58956}{n^{4/3}} + \frac{21.2862}{n^{2}} - \frac{17.6087}{n^{7/3}} + \frac{7.4987}{n^{8/3}} - \frac{33.4939}{n^{3}} \nn\\
& + \frac{48.2321}{n^{10/3}} - \frac{44.7412}{n^{11/3}} + \frac{34.7669}{n^{4}} - \frac{67.6941}{n^{13/3}}\,,\nn\\
 \Delta=\fft{131}{20}\,,\quad {\rm fit}[n]&=-\frac{11.4888}{n} + \frac{4.333}{n^{4/3}} + \frac{56.8047}{n^{2}} - \frac{48.0957}{n^{7/3}} + \frac{22.2311}{n^{8/3}} - \frac{158.142}{n^{3}} \nn\\
& + \frac{227.361}{n^{10/3}} - \frac{228.236}{n^{11/3}} + \frac{307.469}{n^{4}} - \frac{596.036}{n^{13/3}}\,.
\ee

\section{Asymptotic OPE method in position space}
\label{app: OPE x=0}

In this Appendix, we review the asymptotic OPE method in position space \cite{Fitzpatrick:2019zqz} and provide more fits in addition to \eqref{eq: fits x=0 original} and \eqref{eq: fits =0} for other values of $\Delta$.

\subsection{Review asymptotic OPE method in position space}

The key insight of \cite{Fitzpatrick:2019zqz} is to solve \eqref{eq: bulk scalar eq p} perturbatively in $1/r$ that matches with the structure of conformal block expansion, which is \eqref{eq: block expansion T} for thermal CFTs. However, the original set-up in \cite{Fitzpatrick:2019zqz} is to match with heavy-heavy-light-light conformal block expansion in the light-light OPE channel, where in the limit $(\tau,x)\rightarrow0$ with fixed $\eta$ we have
\be
& \langle H|\mathcal{O}(\tau,x)\mathcal{O}(0)|H\rangle:=\langle H(1)\mathcal{O}(\tau,x)\mathcal{O}(0)H(\infty)\rangle=\fft{1}{(\tau^2+x^2)^{\Delta}}\sum_{\Delta',J}C_{\Delta',J}G_{\Delta',J}^{00}\,,\nn\\
&  G_{\Delta',J}^{00}=(\tau^2+x^2)^{\fft{\Delta'}{2}}\fft{(d-2)_J}{\left(\fft{d}{2}-1\right)_J}\hat{C}_J^{\fft{d}{2}-1}\left(\eta\right)\,.\label{eq: block expansion HL}
\ee
It agrees with the expansion with finite temperature \eqref{eq: block expansion T} by identifying 
\be
C_{\Delta',J}=\fft{\left(\fft{d}{2}-1\right)_J}{(d-2)_J} c_{\Delta',J} \langle O'\rangle_\beta
\ee
This is a toy example of thermalization in the heavy limit $\Delta_H\rightarrow\infty$.

In position space, the scalar bulk-to-boundary propagator is
\be
G_{b\partial}^{\rm AdS}(\tau,x)\sim \fft{r^{\Delta}}{(1+r^2(\tau^2+x^2))^{\Delta}}\,.
\ee
The strategy is to define the variables
\be
w^2=1+r^2(\tau^2+x^2)\,,\quad \rho=r x\,.
\ee
Then we solve the equation by the following ansatz asymptotically in $1/r$
\be
\phi(r,w,\rho)=\fft{r^\Delta}{w^{2\Delta}}\Psi(r,w,\rho)\,,\quad \Psi(r,w,\rho)=\sum_{n=0}\fft{\mu^n g_{n}(w,\rho)}{r^{nd}}\,,
\ee
where we keep $(w,\rho)$ finite. Physically, we are probing the OPE limit $(\tau,x)\rightarrow 0$ with fixed $\eta$ by taking the near boundary expansion. In even dimensions, $g_n(w,\rho)$ further simplifies to truncated polynomials
\be
g_{n}(w,\rho)=\sum_{m=0}^{2n} \sum_{k=-2n}^{dn-m}Q_{nmk}  \rho^m w^k\,.
\ee
One can then turn the bulk equation of motion into a recursion equation for $Q_{nmk}$ in the two variables $(m,k)$ , as shown in \cite{Fitzpatrick:2019zqz,Fitzpatrick:2019efk,Li:2019tpf}. In the end, we take $r\rightarrow\infty$ and match with the conformal block expansion \eqref{eq: block expansion HL} to read off the OPE coefficients $C_{\Delta',J}$ or $c_{\Delta',J}$. In the thermal case with $x=0$, the matching procedure simplifies as we take $x=0$ and thus we have \cite{Fitzpatrick:2019zqz}
\be
a_{nd}=Q_{n,0,dn}\pi^{dn}\,.
\ee
The explicit low-lying results can be found in \cite{Fitzpatrick:2019zqz,Fitzpatrick:2019efk,Li:2019tpf}.

We note that the results contain poles for integer $\Delta$ \cite{Fitzpatrick:2019zqz}. This reflects the operator mixing between multi-stress-tensors and double-trace operators, as explored in \cite{Li:2019tpf,Li:2020dqm,Barrat:2025twb}. Nevertheless, since we are interested in the retarded correlator, the corresponding OPE coefficients are multiplied with factors $\sin\left(\fft{\Delta'-2\Delta}{2}\pi\right)$ that cancel the divergence and eliminate the double-trace contributions in the planar limit (because $\Delta'=2\Delta+2\mathbb{Z}$). This was dubbed as the residue relation in \cite{Li:2019tpf,Li:2020dqm}.

\subsection{More fits}

We now provide more fits in $x=0$ case to justify our results \eqref{eq: F fit}, \eqref{eq: fit new formula} and \eqref{eq: Ft fit}.

We find
\be
 \Delta=\fft{91}{20}\,,\quad {\rm fit}^{x=0}[n]&=-0.00176673 - \frac{8.44881}{n} - \frac{3.4789}{n^{4/3}} + \frac{29.279}{n^2}+\cdots\,,\nn\\
\Delta=\fft{111}{20}\,,\quad {\rm fit}^{x=0}[n]&=-0.00379219 - \frac{13.2136}{n} - \frac{5.41094}{n^{4/3}} + \frac{74.8417}{n^2}+\cdots\,,\nn\\
 \Delta=\fft{131}{20}\,,\quad {\rm fit}^{x=0}[n]&=-0.00524339 - \frac{18.965}{n} - \frac{7.68903}{n^{4/3}} + \frac{155.314}{n^2}+\cdots\,,\nn\\
\Delta=\fft{3}{2}\,,\quad {\rm fit}^{x=0}[n]&=0.0167498 - \frac{2.59942 \times 10^{-8}}{n} + \frac{0.179768}{n^{4/3}} - \frac{0.000129345}{n^2}+\cdots\,,\nn\\
 \Delta=\fft{5}{2}\,,\quad {\rm fit}^{x=0}[n]&=0.00658228 - \frac{1.76152}{n} - \frac{0.649186}{n^{4/3}} + \frac{0.7547}{n^2}+\cdots\,,\nn\\
 \Delta=\fft{7}{2}\,,\quad {\rm fit}^{x=0}[n]&=0.00144666 - \frac{4.50651}{n} - \frac{1.83903}{n^{4/3}} + \frac{7.44834}{n^2}+\cdots\,,\nn\\
\Delta=\fft{35}{11}\,,\quad {\rm fit}^{x=0}[n]&=0.00278039 - \frac{3.52631}{n} - \frac{1.42112}{n^{4/3}} + \frac{4.26657}{n^2}+\cdots\,.
\ee
After factorizing out the more accurate leading asymptotic behavior, the fits lead to
\begin{align*}
\Delta = \frac{91}{20} : & \quad \text{fit}^{x=0,\text{new}} = - \frac{8.46376}{n} - \frac{3.48506}{n^{4/3}} + \frac{29.3308}{n^2} + \dots, \\
\Delta = \frac{111}{20} : & \quad \text{fit}^{x=0,\text{new}} = - \frac{13.2639}{n} - \frac{5.43154}{n^{4/3}} + \frac{75.1266}{n^2} + \dots, \\
\Delta = \frac{131}{20} : & \quad \text{fit}^{x=0,\text{new}} = - \frac{19.065}{n} - \frac{7.72955}{n^{4/3}} + \frac{156.133}{n^2} + \dots, \\
\Delta = \frac{3}{2} : & \quad \text{fit}^{x=0,\text{new}} = - \frac{2.5566 \times 10^{-8}}{n} + \frac{0.176807}{n^{4/3}} - \frac{0.000127214}{n^2} + \dots, \\
\Delta = \frac{5}{2} : & \quad \text{fit}^{x=0,\text{new}} = - \frac{1.75}{n} - \frac{0.644941}{n^{4/3}} + \frac{0.749765}{n^2} + \dots, \\
\Delta = \frac{7}{2} : & \quad \text{fit}^{x=0,\text{new}} = - \frac{4.5}{n} - \frac{1.83638}{n^{4/3}} + \frac{7.43758}{n^2} + \dots, \\
\Delta = \frac{35}{11} : & \quad \text{fit}^{x=0,\text{new}} = - \frac{3.51653}{n} - \frac{1.41718}{n^{4/3}} + \frac{4.25474}{n^2} + \dots\,.
\end{align*}
We can see agreement with \eqref{eq: fit new formula} for the relevant terms.

\bibliographystyle{JHEP}
\bibliography{refs}

\end{document}